\documentclass[aps,twocolumn,groupedaddress,showpacs]{revtex4-1}
\usepackage{graphicx,amsfonts,amssymb,amsmath,subfigure,accents}

\usepackage{color}

\begin{document}

\title{Quantum and classical noise characteristics of parametrically driven cavity solitons in dispersive Kerr resonators}

\author{Sophie S.~Shamailov}
\affiliation{Department of Physics, University of Auckland, Auckland, New Zealand.}
\affiliation{The Dodd-Walls Centre for Photonic and Quantum Technologies, Dunedin, New Zealand.}
\author{Miro Erkintalo}
\email{m.erkintalo@auckland.ac.nz}
\affiliation{Department of Physics, University of Auckland, Auckland, New Zealand.}
\affiliation{The Dodd-Walls Centre for Photonic and Quantum Technologies, Dunedin, New Zealand.}

\date{\today}

\begin{abstract}
Temporal cavity solitons generated in monochromatically driven dispersive Kerr resonators offer an attractive avenue for on-chip optical frequency comb generation. Key to many of their applications is to understand how noise --- both technical and quantum --- affects the soliton states, which has accordingly been extensively investigated. Here, we report on a comprehensive theoretical study that elucidates how technical and quantum fluctuations impact a new type of soliton structure that has very recently been predicted and observed in dispersive Kerr resonators under conditions of bichromatic driving: the pure-Kerr parametrically driven cavity soliton (PDCS). We examine how classical laser phase noise transfers from the two pump fields onto the soliton frequency comb, and we calculate the solitons' fundamental quantum-limited timing jitter and two-mode squeezing spectra. In each case, we find that PDCSs can out-perform conventional cavity solitons with comparable characteristics, even when driven by two uncorrelated lasers. Our results demonstrate that pure-Kerr PDCSs can offer unprecedented performance in noise-sensitive photonic applications and as a quantum resource.

\end{abstract}


\maketitle

\emph{Introduction}---Temporal cavity solitons (CSs) are steady-state field configurations in monochromatically driven dispersive Kerr resonators that correspond to pulses of light circulating the resonator indefinitely. Since first discovered in 2010 \cite{FLeo}, they have been extensively studied due to a plethora of promising applications \cite{Kipp}. In particular, CSs generated in monolithic microresonators have been widely recognized as an attractive avenue for generating coherent optical frequency combs in photonic integrated circuits, with demonstrated applications ranging from telecommunications and metrology to sensing and photonic computing.

For many photonic applications, it is crucial to understand the solitons' noise characteristics. To first order, the phase noise of the driving laser is copied by all of the CS comb lines, but higher-order (dispersive or nonlinear) effects can give rise to variations and so-called ``quiet points'' -- particular comb lines with phase noise lower than that of the drive \cite{Torres}. At the limit of zero phase (or other technical) noise, the solitons' noise (timing jitter) is governed by quantum fluctuations, which have been studied analytically \cite{jitter} and probed experimentally \cite{quantum_solitons4}.

Whilst CS frequency combs have hitherto been primarily used in classical photonic applications \cite{Kipp}, recent years have witnessed increasing appreciation that such solitons could also constitute the next milestone in a variety of quantum technological applications \cite{Chembo, soliton_microcombs, quantum_solitons1},  building on top of the large body of work centred around the vacuum state \cite{Kues, Wang, Pfister3, Wu, Kues2022, Vaidya2020, Gaeta2020, Chembo_published, quantum_solitons5, squeezed_microcombs}. Specifically, CS combs could offer a route to creating high-dimensional, strongly entangled and squeezed continuous-variable cluster states that can have potential applications in fields of quantum computation and information, metrology, sensing and communications \cite{Ref32, secrets, Wang2020, ChemboRef49, Ligo2013, Ligo2023, sensing, Spectroscopy, fromStuart, teleportation, Pfister3, Zhang2021, Zhong2020, DenseCoding, Nicolas, Nielsen2006, Wu, Pfister2_Ref25, Pfister, quantumcomp}. For this reason, the quantum optical characteristics of CSs --- e.g., pairwise and multimode squeezing and entanglement, and the two-photon correlation function $g^{(2)}(t)$ --- have been extensively studied over the past decade \cite{Chembo, soliton_microcombs, quantum_solitons1}.

Recently, it has been discovered that dispersive Kerr resonators can support a new type of soliton state under conditions of \emph{bichromatic} driving \cite{PDCS}. In contrast to conventional CSs, which are spectrally centred around the single monochromatic drive, these new soliton states arise from the parametric mixing of the two driving fields and are spectrally centred in between the two inputs. In addition to spectral separation from the inputs, these pure-Kerr \emph{parametrically driven cavity solitons} (PDCSs) exhibit several other interesting features that could be beneficial for distinct photonic and quantum applications~\cite{nico}. However, having only been predicted and observed recently, the comparative characteristics of pure-Kerr PDCSs have yet to be explored. Noise and quantum optical characteristics are of particular interest: on the one hand, the parametric amplification that underpins PDCSs can theoretically offer a sub-classical (squeezed) noise landscape, but on the other hand, the use of two separate driving lasers with uncorrelated (phase) fluctuations can be speculated to degrade the noise characteristics of PDCSs compared to monochromatically driven CSs. A recent study has predicted that pure-Kerr PDCSs can be associated with strong multimode squeezing~\cite{avik}, yet several questions remain around the solitons' noise characteristics.

In this Letter, we report on an extensive analytical and numerical study of the quantum optical characteristics as well as technical (classical) phase noise properties of pure-Kerr PDCSs. We explore how classical phase noise from the driving lasers transfers onto the PDCS comb, analytically evaluate the solitons' fundamental (quantum-limited) timing jitter~\cite{jitter}, and compute the two-mode squeezing spectra associated with the soliton comb lines. We find that, across all the measures considered, PDCSs out-perform conventional monochromatically driven CSs: they exhibit stronger two-mode squeezing (and hence entanglement \cite{Sqzng_Entglmt,Sqzng_Entglmt2}), lower fundamental timing jitter, and despite being driven by two uncorrelated lasers, display lower phase noise for comparable laser noise profiles. Taken together, our results highlight that PDCSs can offer comparative performance advantages in noise-sensitive and quantum applications.

 \begin{figure}[!t]
    \centering
    \includegraphics[width = \columnwidth, clip=true]{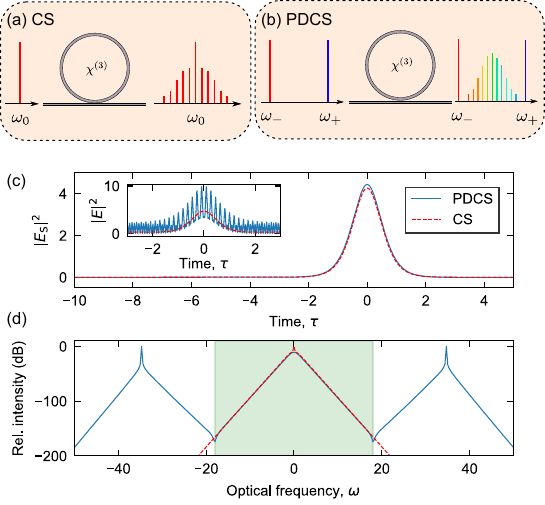}
    \caption{(a) and (b) show schematic illustrations of (a) CS and (b) PDCS generation in a monochromatically and bichromatically driven dispersive $\chi^{(3)}$ Kerr resonator, respectively. (c) and (d) show numerically simulated temporal and spectral profiles, respectively, of a PDCS (solid blue curve) and a CS (red dashed curve). To isolate the soliton component, $E_\mathrm{S}(\tau)$, the constant background was removed from the CS field and a spectral band-pass filter (shaded green region in (d) shows the pass-band; inset in (c) shows the intensity profiles before background removal) was used on the PDCS before obtaining the intensity profiles in (c). PDCS simulations use $d_2=-2$ and $d_4=0.02$, $\Delta_{\pm}=0$, $S_\pm \approx 1.89$, and $\Omega_\mathrm{p} \approx 34.7$, yielding effective parameters $\mu=1.37$ and $\Delta_{\mbox{eff}}=1.2$. CS simulations use $d_2 = -2$, $\Delta_0 = 2.4$ and $S_0 = 1.42$. All $d_k$ that are not quoted are zero (throughout the Letter). } 
    \label{solidemo}
\end{figure}

\emph{Models}---We are interested in dispersive resonators with $\chi^{(3)}$ Kerr nonlinearity that are externally (coherently) driven. Depending on the driving configuration and specific parameters, the same system can support both conventional CSs as well as PDCSs [see Fig.~\ref{solidemo}]. The system is classically modelled by the generalized Lugiato-Lefever equation, which we write in dimensionless form (for normalization, see Supplementary Information):
\begin{align}
\frac{\partial E(t,\tau)}{\partial t} &= \left[-1 + i(|E|^2 -\Delta_0)+ iD\left(i\frac{\partial}{\partial \tau}\right)\right]E \nonumber \\
 &+ S(t,\tau). \label{LLE}
\end{align}
Here, $E(t,\tau)$ describes the complex amplitude of the slowly varying electric field envelope within the resonator, with $t$ and $\tau$ being the slow- and fast-time variables describing the evolution of the field envelope over multiple round trips and its profile over a single round trip, respectively. The terms on the right-hand side of Eq.~\eqref{LLE} from left to right describe losses, the Kerr nonlinearity, detuning between the soliton carrier frequency and the closest cavity resonance, group-velocity dispersion with $D\left(\frac{\partial}{\partial \tau}\right) = \sum_{k\geq2} \frac{d_k}{k!}\frac{\partial^k}{\partial\tau^k}$ the normalized dispersion operator, and coherent driving with $S(t,\tau)$ the normalized driving amplitude. 

For conventional CSs, the frequency of the driving field coincides with the carrier frequency of the solitons, such that $S(t,\tau) = S_0$ with $S_0$ a constant scalar. For PDCSs, the soliton carrier frequency is positioned in between two driving fields, which can be modelled viz. 
\begin{equation}
S(t,\tau) = S_+ e^{-i\Omega_\mathrm{p}\tau + iat} +  S_- e^{i\Omega_\mathrm{p}\tau - iat}.
\end{equation}
Here, $S_\pm$ are the normalized amplitudes of the two driving fields, whilst the parameters $\Omega_\mathrm{p}$ and $a$ describe  the angular frequency separation of the driven resonances from the soliton carrier frequency that lies in between the two pumps. Specifically, $\Omega_\mathrm{p}$ captures the coarse frequency separation as an integer multiple of the resonator free-spectral range, whilst $a = (\Delta_+ - \Delta_- + D(\Omega_\mathrm{p}) - D(-\Omega_\mathrm{p}))/2$ accounts for the detunings $\Delta_\pm$ of the individual pump fields from the resonances they drive. [The pump detunings also define $\Delta_0$ viz.~$\Delta_0 = (\Delta_+ + \Delta_- + D(\Omega_\mathrm{p}) + D(-\Omega_\mathrm{p}))/2$].

Both conventional CSs and PDCSs correspond to steady-state solutions of Eq.~\eqref{LLE}, consisting of a superposition of a bright pulse ($E_\mathrm{S}(\tau)$) and quasi-continuous-wave fields ($E_\mathrm{p}(\tau)$) at the pump frequencies, $E(\tau) = E_\mathrm{S}(\tau) + E_\mathrm{p}(\tau)$. In both cases, the pulse has an approximately hyperbolic secant amplitude profile, with 
\begin{equation}
E_\mathrm{S}(\tau) \approx \sqrt{2}\beta \text{sech}(\beta\tau)e^{i(\phi+\theta)}. \label{CSsol}
\end{equation}
For CSs, $\theta_\mathrm{CS} = \text{arg}[S_0]$, $\cos(\phi_\mathrm{CS}) \approx \sqrt{8\Delta_0}/(\pi S_0)$, and $\beta_\mathrm{CS} = \sqrt{\Delta_0}$, whilst for PDCSs we have $\theta_\mathrm{PDCS} = \text{arg}[\mu]/2$ and $\cos(2\phi_\mathrm{PDCS})=1/|\mu|$, where $\mu = 2iE_+E_-$ is the effective parametric driving strength with $E_\pm$ the complex amplitudes of the intracavity fields at the driving frequencies, and $\beta_\mathrm{PDCS} = \sqrt{\Delta_\mathrm{eff} + |\mu|\sin(2\phi_\mathrm{PDCS})}$, with $\Delta_\mathrm{eff} = \Delta_0 - 2(|E_+|^2 + |E_-|^2)$ an effective detuning.  Within the normalization used, the PDCS existence and properties are primarily governed by the effective parameters $\mu$ and $\Delta_\mathrm{eff}$, whilst conventional CSs are governed by the normalized driving strength $S_0$ and detuning $\Delta_0$. In comparing the noise and quantum optical characteristics of CSs and PDCSs below, we always consider parameter combinations yielding $\beta_\mathrm{CS} \approx \beta_\mathrm{PDCS}$ to ensure similar soliton temporal and spectral characteristics [see e.g. Fig.~\ref{solidemo}]. Moreover, we focus on the case of free-running PDCSs, reserving analyses pertaining to recently discovered self-aligned states~\cite{Moille2026} for future work.

\emph{Phase noise}---We first consider the transfer of classical driving laser phase noise onto the soliton frequency combs. To this end, we perform simulations where driving field phase noise is included by modifying the relevant driving field amplitudes viz.~$S_{0,\pm}\rightarrow S_{0,\pm} \exp[i\delta\theta_{0,\pm}(t)]$, where the phase fluctuations $\delta\theta_{0,\pm}(t)$ correspond to the indefinite integral of uncorrelated Gaussian white noise with zero mean (see Supplementary Information for details). Driving fields of this form, when time-averaged, yield a Lorentzian spectrum with optical linewidth $\Delta f_{0,\pm}$ directly proportional to the variance of the phase fluctuations. To examine the resulting phase fluctuations of the soliton comb, we express the intracavity field in the Fourier domain as $E(t,\tau) = \sum_{l=-K}^K E_l(t) e^{-2\pi il\tau/T}$, where $E_l(t)$ is the modal amplitude of the comb line with relative index $l$ ($l = 0$ corresponds to the soliton carrier), $2K+1$ is the number of modes modeled, and $T$ is the simulated domain in fast time (physically corresponding to the resonator round-trip time). We analyze the phase fluctuations of different comb lines by Fourier transforming the relevant mode amplitudes $E_l(t)$, the absolute value squared of which yields the optical line profile. (To avoid confusion, we use $\omega$ and $\Omega$ as normalized angular frequency Fourier variables corresponding to $\tau$ and $t$, respectively.)

 \begin{figure}[!t]
    \centering
    \includegraphics[width = \columnwidth, clip=true]{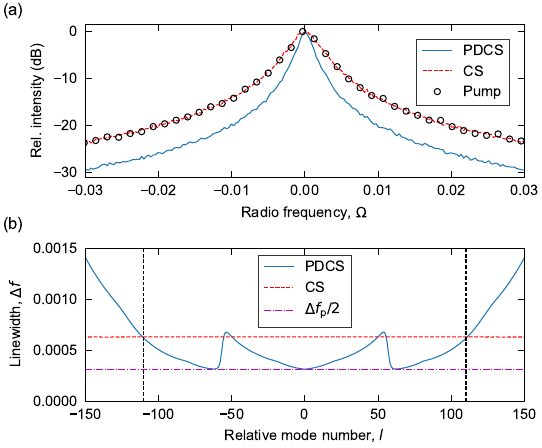}
    \caption{(a) Numerically simulated line profile of the $l=0$ comb mode for a PDCS (solid blue curve) and a corresponding CS (red dashed curve). Black circles show the pump profile used in the simulations. (b) 3-dB widths of the PDCS (solid blue curve) and CS (red dashed curve) comb lines as a function of relative mode number $l$. Vertical black lines indicate the pumped modes in PDCS simulations, and horizontal dash-dotted magenta line indicates the pump linewidth divided by two. All simulation parameters as in Fig.~\ref{solidemo}. The results were obtained by averaging over 500 independent noise realizations.} 
    \label{pumpnoise}
\end{figure}

Figure~\ref{pumpnoise} shows illustrative results from our simulations. Panel (a) compares the line profiles of the $l=0$ mode for conventional CSs (red dashed line), PDCSs (blue solid line), and one of the driving fields (black circles; all driving fields considered are uncorrelated but have the same statistics). The CS line profile follows exactly that of its monochromatic driving field, which is expected given that CSs are known to be phase-locked to their drive [recall that $\theta_\mathrm{CS}=\text{arg}[S_0]$ in Eq.~\eqref{CSsol}]. Interestingly, however, the PDCS line exhibits a clearly narrower width compared to the conventional CS and the driving fields, highlighting lower phase noise. 

The fact that PDCSs can enjoy reduced phase noise compared to CSs despite being driven by two uncorrelated lasers may appear surprising, but can readily be understood from the underlying physics. Parametrically driven cavity solitons are phase-locked to the parametric driving term viz.~$\theta_\mathrm{PDCS} = \text{arg}[\mu(t)]/2 = \text{arg}[iE_+(t)E_-(t)]/2$, with the intracavity fields at the pump frequencies, $E_\pm$, in turn phase-locked to the drives. Thus, as the fields $E_\pm$ inherit the fluctuations of the drives ($E_\pm(t)\propto \exp[i\delta\theta_\pm(t)]$), the PDCS phase fluctuation is $\delta\theta_\mathrm{PDCS} = (\delta\theta_+ + \delta\theta_-)/2$, yielding the variance 
\begin{equation}
    \text{var}(\delta\theta_\mathrm{PDCS}) = \frac{\text{var}(\delta\theta_+) + \text{var}(\delta\theta_-) + 2 \text{cov}(\delta\theta_+,\delta\theta_-)}{4}. \label{PDCSvar}
\end{equation}
For uncorrelated drives that have the same variance, $\text{var}(\delta\theta_+) = \text{var}(\delta\theta_-) \equiv \sigma^2$ and $\text{cov}(\delta\theta_+,\delta\theta_-)=0$, so we have $\text{var}(\delta\theta_\mathrm{PDCS}) = \sigma^2/2$, predicting the soliton linewidth to be half of that of the drives -- exactly as seen in the simulation results in Fig.~\ref{pumpnoise}. For correlated pump beams where $\delta\theta_+ = \delta\theta_-$, $\text{cov}(\delta\theta_+,\delta\theta_-) = \sigma^2$, such that $\text{var}(\delta\theta_\mathrm{PDCS}) = \sigma^2$ (confirmed by numerical simulations). Of course, if the phase fluctuations of the drives are completely anti-uncorrelated, then $\text{cov}(\delta\theta_+,\delta\theta_-) = -\sigma^2$, and we obtain $\text{var}(\delta\theta_\mathrm{PDCS}) = 0$; in this case, the phase fluctuations of the two pumps exactly cancel and the PDCS does not exhibit any (classical) phase noise in the $l=0$ mode, again in agreement with simulations.

To gain more insights, Fig.~\ref{pumpnoise}(b) shows the normalized full-width at half maximum linewidths, $\Delta f$, for CS (red dashed line) and PDCS (blue solid line) comb components as a function of the relative mode index $l$. As can be seen, each component of the CS comb exhibits the same linewidth, which is equal to the linewidth of the pump. (This result is expected given that we are ignoring higher-order dispersion~\cite{Torres}.) In stark contrast, the linewidths of the PDCS comb components vary nontrivially as a function of $l$, though we note that (i) the $l=0$ mode follows the theoretical prediction given by Eq.~\eqref{PDCSvar} and (ii) comb lines at both pump frequencies follow the linewidths of the pumps. 

Extensive simulations reveal that the dependence of the PDCS comb linewidth on $l$ is due to two distinct factors that distinguish the full system (as modeled by Eq.~\eqref{LLE}) from the parametrically driven nonlinear Schr\"odinger equation~\cite{PDCS,nico} that admits exact soliton solutions of the form given by Eq.~\eqref{CSsol}. First, the nonlinear interaction between the PDCS and the intracavity fields at the pump frequencies gives rise to frequency combs around both pumps. The mixing of theses secondary combs will provide additional pathways to driving the PDCS comb, contributing to the phase fluctuations of that comb. In simulations that suppress (via numerical filtering) the formation of sub-combs around the pump frequencies, the dependence of comb linewidths on $l$ disappears, with all lines (except at the pump frequencies) obeying the prediction given by Eq.~\eqref{PDCSvar}. The second factor that contributes to the dependence of the PDCS comb linewidth on $l$ is the uncorrelated character of the pump fluctuations used to obtain the simulation results in Fig.~\ref{pumpnoise}. Indeed, the dependence is not observed in simulations with fully correlated pump fluctuations (even when sub-combs around both pumps are allowed to grow). Taken together, these findings suggest that the non-trivial dependence of the PDCS comb linewidths is driven by the uncorrelated fluctuations of the sub-combs around the pumps. 

 \begin{figure}[!t]
    \centering
    \includegraphics[width = \columnwidth, clip=true]{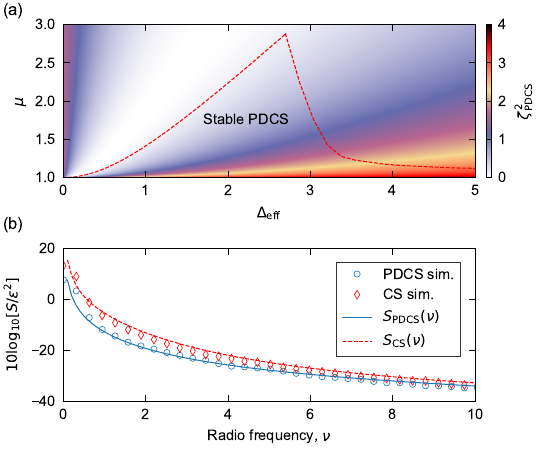}
    \caption{(a) Coefficient $\zeta_\mathrm{PDCS}^2$ appearing in Eq.~\eqref{CS_jitter_formula_ms} across the ($\Delta_\mathrm{eff},\mu$) parameter space. Values less than four signal PDCSs to exhibit lower jitter than corresponding CSs. Dashed red curve delineates the region where stable PDCSs exist. (b) Solid blue (dashed red) curves show PDCS (CS) timing jitter spectra calculated from Eq.~\eqref{CS_jitter_formula_ms} with $\Delta_\mathrm{eff} = 2.5$ and $\mu = 2.5$ ($\Delta_0 = 4$, $S_0 = 2$). Blue circles (red diamonds) show corresponding results from numerical simulations with dispersion parameters as in Fig.~\ref{solidemo}, with $\Delta_\pm = 0$, $S_\pm \approx 4.34$, and $\Omega_\mathrm{p} \approx 34.8$. The simulated spectra were averaged over 500 independent noise realizations and used $\varepsilon = \sqrt{\gamma L \hbar \omega_0/(\alpha t_\mathrm{R})} = 0.001$ --- an exaggerated value to expedite the simulations (the noise spectrum simply scales with $\varepsilon^2$).} 
    \label{jitter_fig}
\end{figure}

\emph{Fundamental timing jitter}---Despite the non-trivial dependence of the PDCS comb noise on the mode order $l$, the above results overall show that PDCSs can experience lower technical phase noise than the drives and thus lower phase noise than corresponding CSs driven with a single monochromatic input. In the limit where the drives exhibit negligible classical noise, the fluctuations of both types of solitons is governed by fundamental quantum fluctuations. To understand how the quantum-limited noise of PDCSs compares with that of conventional CSs, we calculated the fundamental timing jitter in the PDCS repetition rate following the method used in~\cite{jitter} for CSs. We find that both types of solitons are associated with a qualitatively similar fundamental timing jitter spectrum, described in normalized units by the following analytical formula (the full details of the calculation as well as the corresponding formula in dimensional units are provided in the Supplementary Information):
\begin{equation}
\label{CS_jitter_formula_ms}
S_\mathrm{CS,PDCS}(\Omega) = \frac{\gamma L \hbar \omega_0}{\alpha t_R} \frac{2}{3\mathcal{E}\Omega^2}\left(\frac{\zeta_\text{CS,PDCS}^2}{w_{ss}^2(4+\Omega^2)} + \frac{\pi^2 w_{ss}^2}{4}\right),
\end{equation}
where $\gamma$ is the Kerr nonlinearity coefficient, $L$ is the length of the resonator, $\omega_0$ is the soliton carrier frequency (average of the two driving frequencies for PDCSs), $t_R$ the round-trip time, $\alpha=\pi/\mathcal{F}$ where $\mathcal{F}$ is the finesse, $A_{ss}$, $w_{ss}$, $\phi_{ss}$ are the (dimensionless) amplitude, width and phase of the soliton, respectively, and $\mathcal{E}=2A_{ss}^2w_{ss}$ is the soliton energy. 

For conventional CSs, the coefficient $\zeta_\mathrm{CS} = -2$ in Eq.~\eqref{CS_jitter_formula_ms}, and the expression agrees with the result obtained in \cite{jitter}. For PDCSs, the coefficient
\begin{equation}
    \zeta_\mathrm{PDCS} = -2 - \frac{2}{3}\pi^2 w_{ss}^2 m,
\end{equation} 
where $m =\mathcal{R}(E_+ E_-)\cos(2\phi_{ss}) + \mathcal{I}(E_+ E_-)\sin(2\phi_{ss})$ with $\mathcal{R}(\cdot)$ and $\mathcal{I}(\cdot)$ denoting the real and imaginary parts, respectively. The pseudo-colour plot in Fig.~\ref{jitter_fig}(a) shows $\zeta^2_\mathrm{PDCS}$  across the effective parameter space ($\Delta_\mathrm{eff},\mu$), obtained via direct computation. As can be seen, we find that $\zeta^2_\mathrm{PDCS} < 4$ at all points, which implies that $S_\mathrm{PDCS}(\Omega) < S_\mathrm{CS}(\Omega)$ for all $\Omega\neq 0$. This result thus theoretically predicts that PDCSs exhibit lower fundamental timing jitter compared to ordinary CSs.

To test the theoretical prediction described above, we performed stochastic simulations where random (Gaussian) noise fluctuations are added to Eq.~\eqref{LLE} --- see Supplementary Information for details --- and we extracted the jitter in the soliton's temporal center of mass. (The absolute value of the Fourier transform of the jitter time series yields a numerical simulation of the corresponding jitter spectrum.) The blue circles and red diamonds in Fig.~\ref{jitter_fig}(b) show the PDCS and CS jitter spectra as obtained from our simulations, respectively. Also shown are the corresponding analytical predictions given by Eq.~\eqref{CS_jitter_formula_ms} as blue solid ($S_\mathrm{PDCS}(\Omega)$) and red dashed ($S_\mathrm{CS}(\Omega)$) lines. As can be seen, the numerical simulation results agree with the analytical prediction given by Eq.~\eqref{CS_jitter_formula_ms}, thus also confirming that PDCSs indeed exhibit lower fundamental timing jitter compared to ordinary CSs.

\emph{Two-mode squeezing}---We finally analyze the two-mode squeezing of PDCSs in comparison with conventional CSs. To this end, we employ a mathematical framework that is analogous to that published in~\cite{Chembo} and which is written out in detail in the Supplementary Information. The primary observable of interest is the two-mode quadrature spectrum $S_{\phi,l}(\Omega)$, which describes the noise power spectrum of the joint quadrature rotated by angle $\phi$ for comb lines with relative mode number $\pm l$; it is the frequency-resolved variance of the field quadrature (defined by the mixing angle $\phi$) it corresponds to. 

 \begin{figure}[!t]
    \centering
    \includegraphics[width = \columnwidth, clip=true]{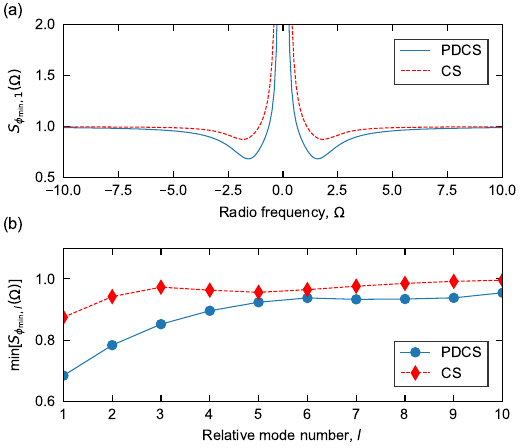}
    \caption{(a) Two-mode quadrature spectra for mode pair $l=1$ for a PDCS (solid blue curve) and corresponding CS (dashed red curve). Both spectra are calculated for their respective optimal mixing angle $\phi_\mathrm{min}$: for the PDCS case, $\phi_\mathrm{min} = 1.26$; for the CS case, $\phi_\mathrm{min} = 2.18$. (b) The minimal value attained by the two-mode quadrature spectra at the optimal mixing angle $\phi_\mathrm{min}$ for a PDCS (blue circles) and a corresponding CS (red diamonds), as a function of mode pair number. Solid and dashed curves are guides to the eye. Parameters as in Fig.~\ref{solidemo} with critical coupling assumed (squeezing depends on the coupling conditions -- see Supplementary Information).} 
    \label{QuadSpecsFig}
\end{figure}

We are interested in the maximal two-mode squeezing attainable, and therefore present results for mixing angles $\phi_\mathrm{min}$ that provide the smallest values for $S_{\phi,l}(\Omega)$; values below unity signal squeezing and the smaller this value the stronger the squeezing. Figure~\ref{QuadSpecsFig}(a) shows typical results, comparing the quadrature spectra for $l=1$ along the optimal angle $\phi_\mathrm{min}$ for a PDCS and a CS with similar characteristics (see caption for parameters; note that the optimal mixing angle $\phi_\mathrm{min}$ is different for the two cases). As can clearly be seen, the PDCS exhibits noticeably stronger squeezing than the corresponding CS. We repeat the calculations for different mode pairs $\pm l$, and plot in Fig.~\ref{QuadSpecsFig}(b) the minimum value of the quadrature spectra obtained, finding that (i) PDCSs exhibit stronger squeezing than CSs for all $l$, but (ii) the level of squeezing reduces with $l$.

The results shown in Fig.~\ref{QuadSpecsFig} pertain to just one set of parameters, yet convey a general trend. We have performed calculations across a broad range of parameters, finding that PDCSs consistently exhibit stronger squeezing than corresponding CSs. Because the same set of effective PDCS parameters $(\Delta_\mathrm{eff},\mu)$ can be obtained with multiple combinations of physical control parameters (e.g. pump powers and detunings), each of which yields slightly different quadrature spectra, it is not readily possible to map the degree of squeezing into a low-level parameter space. Nonetheless, our calculations show that the level of PDCS squeezing generally reduces (increases) as the effective deutuning $\Delta_\mathrm{eff}$ (the parametric driving strength $\mu$) increases. We also find that, as for conventional CSs, the level of PDCS squeezing is increased under conditions of overcoupling.

\emph{Discussion}---To conclude, we have analytically and numerically studied the noise characteristics of PDCSs in bichromatically driven Kerr resonators. We have analyzed how phase noise from the pump fields transfers onto the solitons and evaluated their fundamental timing jitter and two-mode quadrature spectra. For each of the characteristics considered, we have shown that PDCSs outperform conventional CSs. Our results highlight that PDCSs are fundamentally different from conventional CSs, that they can be particularly attractive for applications such as metrology that benefit from low-noise operation, and well-suited to act as a multi-mode quantum resource for a variety of quantum technologies. 

\emph{Acknowledgments}---We acknowledge useful discussions with Avik Dutt, Rafael Romero Mendez, Sashank Kaushik Sridhar, Samyak Gothi, Gr\'egory Moille, and Kartik Srinivasan. We acknowledge financial support from the Marsden Fund of the Royal Society Te Ap\=arangi of New Zealand and the Dodd-Walls Centre for Photonic and Quantum technologies through the Quantum Technologies Aotearoa programme. 

\bibliography{Refs}

\clearpage

\section*{Supplementary Information}

\section{Methods}
\label{Methods}
Here we provide all the technical details needed to reproduce the results presented in the main text. This includes the equations describing the system, derivations, descriptions of simulations, mathematical definitions of parameters, variables and observables, etc. Please note that in this document, we use self-consistent notation, with the general convention that dimensionless quantities are denoted by the same symbol as their counterparts in physical units, except with a tilde. In the main text, however, we present dimensionless equations but the tildes are dropped for simplicity.
\subsection{Classical equations}
\label{Classical}
\subsubsection{Lugiato-Lefevre equations}
\label{LLEs}
We begin this section by recalling the classical equations which govern the dynamics of the parametrically driven cavity soliton (PDCS; derived in \cite{PDCS}) in a bichromatically driven Kerr resonator, the simplified model they reduce to under reasonable assumptions corresponding to an optical parametric oscillator (likewise originally done in \cite{PDCS}), and for completeness, we also cover the case of a monochromatically driven Kerr cavity which supports the standard cavity soliton (CS).

The main system of interest corresponds to a ring resonator with a $\chi^{(3)}$ Kerr nonlinearity, which is coherently pumped with two continuous wave (CW) beams, and experiences a certain level of dissipation. The dispersion is normal at the pump frequencies but anomalous at the average of these two frequencies (the PDCS forms precisely at this latter frequency). The driving beams pump two resonances of the cavity (possibly with detunings), while the light which eventually builds up and forms the PDCS is supported by a cavity resonance which lies half way between the two driven ones.

The Lugiato-Lefevre equation (LLE) governing this physical system is given by
\begin{eqnarray}
\label{biLLE}
&& \frac{\partial E}{\partial t} = -\frac{\alpha}{t_R} E - i\frac{\delta_0}{t_R} E + i \frac{L}{t_R} \sum\limits_{n\geq2} \frac{\beta_n}{n!} i^n \frac{\partial^n E}{\partial\tau^n} + i\frac{\gamma L}{t_R}|E|^2E\nonumber\\
&& + \frac{\sqrt{\tilde{\theta}}}{t_R} E_{in,+} e^{ibt/t_R}e^{-i\Omega_p\tau} + \frac{\sqrt{\tilde{\theta}}}{t_R} E_{in,-}e^{-ibt/t_R}e^{i\Omega_p\tau}.\ \ \ \
\end{eqnarray}
In this equation, $E(t,\tau)$ is the electric field envelope which depends on slow time $t$ and fast time $\tau$. The loss rate is given by $\alpha=\pi/\mathcal{F}$ where $\mathcal{F}$ is the finesse, $t_R$ is the round-trip time of the light around the cavity (at the central frequency where the soliton forms), $\delta_0$ is a phase detuning an equation for which will be given shortly, $L$ is the length (circumference) of the resonator, $\beta_n$ are dispersion coefficients (the required conditions for these will be stated later), $\gamma$ is the nonlinearity strength, $\tilde{\theta}$ is a dimensionless power coupling coefficient for the driving beams, $|E_{in,\pm}|^2$ are the powers of the two pump beams, and $b$ is a parameter which -- like $\delta_0$ -- depends on the phase detunings of the two driving lasers $\delta_{\pm}$, on the dispersion coefficients, and on the frequency $\Omega_p$ which must be an integer multiple of the cavity free spectral range (FSR; twice this integer is the number of resonances the driven peaks are apart).

Throughout the Supplementary Information, we use the symbol $\omega$ ($\Omega$) for optical (radio) frequencies, i.e.~to denote the frequency variable which is associated with fast (slow) time, $\tau$ ($t$).

Defining the dispersion function as
\begin{equation}
D(\omega) = \sum\limits_{n\geq2} \frac{\beta_n}{n!} \omega^n,
\end{equation}
the two parameters $\delta_0$ and $b$ appearing above are given by
\begin{eqnarray}
\delta_0 &=& \frac{\delta_+ + \delta_- + LD(\Omega_p) + LD(-\Omega_p)}{2},\\
b &=& \frac{\delta_+ - \delta_- + LD(\Omega_p) - LD(-\Omega_p)}{2}.
\end{eqnarray}
In order for the dispersion $D(\omega)$ to have the appropriate curvatures at $\omega=0$ (the frequency on which the PDCS spectrum is centred) and at $\omega=\pm\Omega_p$ (the frequencies of the pump beams in this frequency reference frame), we must have $\beta_2<0$, $\beta_4>0$ and small, while $\beta_3>0$ and an arbitrary $\beta_1$ are optional.

Next, it is necessary to take units out of the LLE (\ref{biLLE}) and to cast it in terms of dimensionless quantities, so that we can meaningfully simulate our system. Define
\begin{eqnarray}
\label{DimlessVars}
\Delta_0 &=& \delta_0/\alpha,\nonumber\\
S_{\pm} &=& E_{in,\pm}\sqrt{\frac{\gamma L \tilde{\theta}_{\pm}}{\alpha^3}},\nonumber\\
\tau_s &=& \sqrt{\frac{|\beta_2|L}{2\alpha}},\nonumber\\
\tilde{t} &=& \alpha t/t_R,\nonumber\\
\tilde{\tau} &=& \tau/\tau_s,\nonumber\\
\tilde{E} &=& \sqrt{\frac{\gamma L}{\alpha}} E,\nonumber\\
d_k &=& \frac{\beta_k L}{\alpha \tau_s^k},\nonumber\\
\tilde{\Omega}_p &=& \tau_s \Omega_p,\nonumber\\
a &=& b/\alpha.
\end{eqnarray}
Thus, the dimensionless LLE becomes
\begin{eqnarray}
\label{dimless_LLE}
&& \frac{\partial \tilde{E}}{\partial \tilde{t}} = -\tilde{E} - i\Delta_0\tilde{E} + i \sum\limits_{n\geq2} \frac{d_n}{n!} i^n \frac{\partial^n\tilde{E}}{\partial\tilde{\tau}^n}\nonumber\\
&& + S_+ e^{ia\tilde{t}}e^{-i\tilde{\Omega}_p\tilde{\tau}} + S_- e^{-ia\tilde{t}}e^{i\tilde{\Omega}_p\tilde{\tau}} + i|\tilde{E}|^2\tilde{E}.
\end{eqnarray}

It is possible to readily establish a link between this model and the system studied in \cite{chi2}, as is done in \cite{PDCS}. To this end, substitute
\begin{equation}
\label{ansatz}
\tilde{E}(\tilde{t},\tilde{\tau}) = \tilde{E}_S(\tilde{t},\tilde{\tau}) + \tilde{E}_+ e^{ia\tilde{t}}e^{-i\tilde{\Omega}_p\tilde{\tau}} + \tilde{E}_- e^{-ia\tilde{t}}e^{i\tilde{\Omega}_p\tilde{\tau}}
\end{equation}
into the dimensionless LLE ($\tilde{E}_{\pm}$ are assumed to be constants), and obtain
\begin{eqnarray}
\label{simple_LLE}
\frac{\partial \tilde{E}_S}{\partial \tilde{t}} &=& -\tilde{E}_S - i\Delta_{\mbox{eff}}\tilde{E}_S + i \sum\limits_{n\geq2} \frac{d_n}{n!} i^n \frac{\partial^n\tilde{E}_S}{\partial\tilde{\tau}^n}\nonumber\\
&+& i|\tilde{E}_S|^2\tilde{E}_S + 2i\tilde{E}_+\tilde{E}_-\tilde{E}_S^{\ast},
\end{eqnarray}
where $\Delta_{\mbox{eff}} = \Delta_0 - 2(Y_+ + Y_-)$ and $Y_{\pm} = |\tilde{E}_{\pm}|^2$. Comparing to Eq.~(2) of \cite{chi2}, we identify $\mu=2i\tilde{E}_+\tilde{E}_-$, which renders the two equations identical, except that dispersion is cut off at second order in \cite{chi2}. In the latter case, the LLE admits an analytical soliton solution, given by
\begin{equation}
\label{analPDCS}
\tilde{E}_S(\tilde{\tau}) = \sqrt{2}\beta \mbox{sech}(\beta\tilde{\tau}) e^{i(\bar{\phi}+\bar{\theta})},
\end{equation}
where
\begin{eqnarray}
\label{phi_eqn}
\cos(2\bar{\phi}) &=& 1/|\mu|,\\
\label{beta_eqn}
\beta &=& \sqrt{\Delta_{\mbox{eff}} + |\mu|\sin(2\bar{\phi})},\\
\label{theta_eqn}
\bar{\theta} &=& \frac{1}{2}\mbox{arg}(2i\tilde{E}_+\tilde{E}_-).
\end{eqnarray}
The parameter space in the $\mu$-$\Delta_{\mbox{eff}}$ plane where solitons (\ref{analPDCS}), solutions of (\ref{simple_LLE}), exist and are stable is shown in Figure 2(a) of Ref.~\cite{chi2}. For each point on this phase diagram, one can search for corresponding solutions of the full model, Eq.~(\ref{dimless_LLE}). In order to find such a full-model PDCS solution, it is necessary (but not sufficient -- see below) for the parameters to satisfy a set of conditions. There is considerable remaining freedom of choice in these conditions, so typically, one can find many different full-model PDCSs that all reduce to the same simple-model PDCS. The process to correctly choose and then calculate parameters is as follows.

After we choose a point ($\mu$, $\Delta_{\mbox{eff}}$) where simple-model PDCSs are stable, we calculate $\bar{\phi}$ from (\ref{phi_eqn}) and $\beta$ from (\ref{beta_eqn}). Then we choose the ratio $Y_+/Y_-$ and compute $Y_{\pm}$ from $|\mu| = 2\sqrt{Y_+ Y_-}$. Next, $\Delta_0$ is determined from $\Delta_{\mbox{eff}} = \Delta_0 - 2(Y_+ + Y_-)$, while the two detunings $\Delta_{\pm}$ are chosen freely, as are the dispersion coefficients, $d_n$. The frequency $\tilde{\Omega}_p$ must be solved for so as to satisfy the equation
\begin{equation}
\Delta_0 = \frac{\Delta_+ + \Delta_- + \tilde{D}(\tilde{\Omega}_p) + \tilde{D}(-\tilde{\Omega}_p)}{2},
\end{equation}
where
\begin{equation}
\tilde{D}(\tilde{\omega}) = \sum\limits_{n\geq2} \frac{d_n}{n!} \tilde{\omega}^n.
\end{equation}
The frequency $a$ can then be calculated according to
\begin{equation}
a = \frac{\Delta_+ - \Delta_- + \tilde{D}(\tilde{\Omega}_p) - \tilde{D}(-\tilde{\Omega}_p)}{2}.
\end{equation}
Once $\tilde{\Omega}_p$ is known, the size of the computational window in $\tilde{\tau}$ -- let us call it $\tilde{T}$ -- must be set to an integer multiple of $2\pi/\tilde{\Omega}_p$. Clearly the integer determining $\tilde{T}$ must be sufficiently large to ensure that the analytical solution (\ref{analPDCS}) comfortably fits in to $[-\tilde{T}/2,\tilde{T}/2]$, and the density of points must be sufficiently high to allow the corresponding frequency spectrum to decay fully (down to the noise level). However, satisfying these requirements on $\tilde{T}$ is not enough to ensure one can find a stable PDCS solution to the full model (\ref{dimless_LLE}). Often, $\tilde{T}$ needs to be much larger than the simple model (\ref{simple_LLE}) suggests. This is due to the fact that the full model includes oscillations of the background far away from the soliton (in the simple model, the intensity vanishes far from the soliton), and these tails interact due to the periodic boundary conditions (as we are modelling a ring resonator). Indeed, depending on the particular choice of full-model parameters, $\tilde{T}$ may need to differ between different PDCSs which all reduce to the same analytical (\ref{analPDCS}). In practice, we determine the required value of $\tilde{T}$ by trial, gradually increasing it until a stable soliton is formed. Note that this is not always possible. %Details on the stability and region of existence of the full-model PDCSs is currently being carefully studied in our group and will be published separately.

Let us return to computing the remaining parameters of the full model. Define $X_{\pm}=|S_{\pm}|^2$, and assume without loss of generality that $S_{\pm}$ are real. The driving strengths are calculated from
\begin{equation}
X_{\pm} = [1+(Y_{\pm} + 2Y_{\mp} - \Delta_{\pm})^2]Y_{\pm}.
\end{equation}
It is then possible to determine
\begin{equation}
\tilde{E}_{\pm} = \frac{S_{\pm}}{1-i(Y_{\pm} + 2Y_{\mp} - \Delta_{\pm})},
\end{equation}
and from here, $\bar{\theta}$ is found via (\ref{theta_eqn}). With the parameters determined, the LLE (\ref{dimless_LLE}) is initialised at the ansatz (\ref{ansatz}) and integrated in $\tilde{t}$ until steady state is reached [we use the split-step Fourier method (SSFM) to solve the LLE]. Note that convergence in the integration step needs to be ensured, as is convergence with the number of points $\tilde{\tau}$ is discretised in.

Finally, for completeness, we write down the LLE for a monochromatically driven Kerr resonator (in dimensionless units):
\begin{equation}
\label{mono_LLE}
\frac{\partial \tilde{E}}{\partial \tilde{t}} = -\tilde{E} - i\Delta_0\tilde{E} + i \sum\limits_{n\geq2} \frac{d_n}{n!} i^n \frac{\partial^n\tilde{E}}{\partial\tilde{\tau}^n} + i|\tilde{E}|^2\tilde{E} + S_0.
\end{equation}
In the above, $\Delta_0=\delta/\alpha$ is the phase detuning of the drive with respect to the nearest cavity resonance, and $S_0$ is the dimensionless driving strength. While there is no exact analytical CS solution to the monochromatically driven LLE, an approximate expression is known and can be used as an initial condition to converge to the exact solution:
\begin{equation}
\label{CS_roughsoln}
\tilde{E} = \tilde{E}_p + \sqrt{2\Delta_0} \mbox{sech}(\sqrt{\Delta_0}\tilde{\tau})e^{i(\bar{\phi}+\bar{\theta})}.
\end{equation}
In the above, $\tilde{E}_p$ is the homogeneous solution to the LLE (lower branch) at the given parameters and $\cos(\bar{\phi})=\sqrt{8\Delta_0}/(\pi S_0)$, $\bar{\theta}=\mbox{arg}(S_0)$. If starting from this approximate expression does not lead to convergence to a standard CS, one can lock on to the soliton for different parameters and follow the solution, gradually changing parameters until one gets to the values of interest, and always using the solution at the previous set of parameters to initialise the solver at the next. Alternatively, one can try a detuning scan, a method to excite solitons starting from an empty cavity. For the monochromatically driven LLE, this is a well-established technique (e.g.~\cite{scan_eg}).

Note that Eq.~(1) in the main manuscript is a generalised LLE which reduces to either (\ref{mono_LLE}) or (\ref{dimless_LLE}), as explained in the main text. Recall that the tildes have been dropped in the main text for simplicity. Likewise, the main text generalizes the soliton solution of both equations into one expression, $E(\tau) = E_S(\tau) + E_p(\tau)$ [see just above Eq.~(3) of the main text], which reduces to Eqs.~(\ref{ansatz}) and (\ref{CS_roughsoln}) with the appropriate substitutions detailed in the main manuscript. Observe that the bars over the angles -- introduced here to differentiate these phases from the angles $\phi$ and $\theta$ used in other contexts in this document -- have been dropped in the main text for simplicity.
\subsubsection{Mode equations}
\label{ModeEqns}
Having familiarised ourselves with the spatio-temporal classical description of the system of interest, we will now follow Ref.~\cite{Chembo} to derive the mode equations -- a system of equations which is connected to the LLE via a Fourier transform and is fully equivalent to it in terms of the physics it contains -- which are more suitable for quantisation later on. We will do this for the bichromatically driven Kerr resonator, while the monochromatic case is directly handled in \cite{Chembo} and can also be obtained as a special case of the bichromatic model.

Recall that our system is driven by two CW fields, at frequencies $\omega_{\pm}$. Let us say that these two driving frequencies are respectively closest to two cavity resonances, with numbers $\ell_{\pm}$. Furthermore, we can define the average of $\omega_{\pm}$ as $\omega_0$, and label the cavity resonance closest to $\omega_0$ by the integer $\ell_0$. If we denote the frequencies of the cavity resonances with numbers $\ell_0$, $\ell_+$ and $\ell_-$ respectively as $\omega_{\ell_0}$, $\omega_{\ell_+}$ and $\omega_{\ell_-}$, then $\omega_0 = \omega_{\ell_0} + \sigma_0$ and $\omega_{\pm} = \omega_{\ell_{\pm}} + \sigma_{\pm}$, where the $\sigma$'s are frequency detunings.

We can expand the cavity resonant frequencies as a Taylor series around the $\ell_0^{\mbox{th}}$ peak:
\begin{equation}
\label{Taylor}
\omega_{\ell} = \omega_{\ell_0} + \sum\limits_{n}\frac{\zeta_n}{n!}(\ell-\ell_0)^n.
\end{equation}
Note that first order in $n$ is also included in this sum, and the series can be truncated at whatever order one likes. The FSR (measured around $\omega_{\ell_0}$) is of course none other than $\zeta_1/(2\pi)$. The intra-cavity electric field can be expanded as
\begin{equation}
\vec{E}(\vec{r},t) = \sqrt{2\frac{\hbar\omega_{\ell_0}}{\epsilon_0 n_{\ell_0}^2}}\sum\limits_{\ell}\frac{1}{2}A_{\ell}(t)e^{-i\omega_{\ell}t}\vec{\Upsilon}_{\ell}(\vec{r}) + \mbox{c.c.},
\end{equation}
where an overarrow denotes a vector in three spatial dimensions, and ``c.c.'' stands for the complex conjugate. In this expansion, $A_{\ell}(t)$ is the dimensionless amplitude of mode $\ell$ which only depends on slow time, scaled such that $|A_{\ell}|^2$ is the photon number. Moreover, $\vec{\Upsilon}_{\ell}(\vec{r})$ is a spatial mode function, $\epsilon_0$ is the permittivity of free space, and $n_{\ell_0}$ is the refractive index at $\omega_{\ell_0}$.

Notice that the frequencies $\omega_{\ell}$ are at this stage unevenly spaced due to dispersion. It is convenient to change to a frame of reference where the envelopes rotate at integer multiples of the FSR rather than the actual frequencies of the cavity modes:
\begin{eqnarray}
&& \vec{E}(\vec{r},t) = \sqrt{2\frac{\hbar\omega_{\ell_0}}{\epsilon_0 n_{\ell_0}^2}}\sum\limits_{\ell}\frac{1}{2}A_{\ell}(t)\times\nonumber\\
&& \times \exp[-i(\omega_{\ell_0} + \zeta_1 (\ell-\ell_0))t]\vec{\Upsilon}_{\ell}(\vec{r})+ \mbox{c.c.}.
\end{eqnarray}
Analogously to the monochromatic case studied in \cite{Chembo}, the mode amplitudes obey
\begin{eqnarray}
&& \frac{dA_{\ell}}{dt} = -i \sum\limits_{n\geq2} \frac{\zeta_n}{n!}(\ell-\ell_0)^n -\frac{1}{2}\Delta\omega_{tot,\ell}A_{\ell} \nonumber\\
&& + \frac{1}{2} \Delta\omega_{tot,\ell_+}\mathcal{F}_{\ell_+}e^{-i [\omega_+ - \omega_{\ell_0} - \zeta_1(\ell_+ - \ell_0)] t}\delta_{\ell,\ell_+} \nonumber\\
&& + \frac{1}{2} \Delta\omega_{tot,\ell_-}\mathcal{F}_{\ell_-}e^{-i [\omega_- - \omega_{\ell_0} - \zeta_1(\ell_- - \ell_0)] t}\delta_{\ell,\ell_-}\nonumber\\
&& -ig_0\sum\limits_{\ell_m,\ell_n,\ell_p} A_{\ell_m}A^{\ast}_{\ell_n}A_{\ell_p} e^{-i(\omega_{\ell_m} - \omega_{\ell_n} + \omega_{\ell_p} - \omega_{\ell})t}\nonumber\\
\label{DrivingTerms}
&& \times\Lambda_{\ell}^{\ell_m,\ell_n,\ell_p}\delta_{\ell_m+\ell_p,\ell_n+\ell},
\end{eqnarray}
where $\Delta\omega_{tot,\ell} = \Delta\omega_{int,\ell} + \Delta\omega_{ext,\ell}$ is the total loss rate of photons from mode $\ell$, made up of irreversible loss of energy to the environment ($\Delta\omega_{int,\ell}$) and intentional coupling of light in and out of the resonator via a fibre ($\Delta\omega_{ext,\ell}$). Furthermore, $\mathcal{F}_{\ell_{\pm}} = \sqrt{\frac{4\Delta\omega_{ext,\ell_{\pm}}}{\Delta\omega_{tot,\ell_{\pm}}^2}}\sqrt{\frac{P_{\pm}}{\hbar\omega_{\pm}}}$, where $P_{\pm}$ are the powers of the driving beams and $\omega_{\pm}$ are their frequencies. In addition, $\delta_{m,n}$ is the Kronecker delta function, and $g_0=\frac{n_2 c \hbar \omega_{\ell_0}^2}{n_{\ell_0}^2V_{eff}}$ where $n_2$ is the nonlinear refractive index, $c$ the speed of light, and $V_{eff}$ is the effective mode volume. Finally, $\Lambda_{\ell}^{\ell_m,\ell_n,\ell_p}$ accounts for coupling between the modes due to spatial overlap.

Define $\sigma_0 = -\delta_0/t_R$, $\sigma_{\pm} = -\delta_{\pm}/t_R$, and $b_0 = -b/t_R$. Given the definitions of $\sigma_{\pm}$, together with the Taylor series (\ref{Taylor}), the frequencies of the driving terms in (\ref{DrivingTerms}) are
\begin{equation}
\omega_{\pm} - \omega_{\ell_0} - \zeta_1(\ell_{\pm} - \ell_0) = \sum\limits_{n\geq2} \frac{\zeta_n}{n!}(\ell_{\pm}-\ell_0)^n + \sigma_{\pm}.
\end{equation}
In accordance with the relation of $\sigma_0$ to $\delta_0$ and $b_0$ to $b$, they are given by
\begin{eqnarray}
\sigma_0 &=& \frac{1}{2}\left[\sigma_+ + \sigma_- + \sum\limits_{n\geq2} \frac{\zeta_n}{n!}(\ell_{+}-\ell_0)^n + \sum\limits_{n\geq2} \frac{\zeta_n}{n!}(\ell_{-}-\ell_0)^n  \right],\nonumber\\
b_0 &=& \frac{1}{2}\left[\sigma_+ - \sigma_- + \sum\limits_{n\geq2} \frac{\zeta_n}{n!}(\ell_{+}-\ell_0)^n - \sum\limits_{n\geq2} \frac{\zeta_n}{n!}(\ell_{-}-\ell_0)^n  \right].\nonumber
\end{eqnarray}
After a little algebra, we have
\begin{equation}
\omega_{\pm} - \omega_{\ell_0} - \zeta_1(\ell_{\pm} - \ell_0) - \sigma_0 = \pm b_0.
\end{equation}

We now change the zero point of frequency and shift it to $\omega_0$ (the average of the driving frequencies) instead of $\omega_{\ell_0}$. We also relabel the resonance peaks such that $\ell_0\rightarrow0$ and all other modes are counted from it (i.e.~$l=\ell-\ell_0$ is the new mode numbering index). The effect of the former is to introduce a detuning term into the right-hand side (RHS) of the mode equations. Overall, (\ref{DrivingTerms}) becomes
\begin{eqnarray}
\label{dimfulMode}
&& \dot{A}_l = -k A_l + i\sigma_0 A_l - i \sum\limits_{n\geq2} \frac{\zeta_n}{n!} l^n A_l + \delta_{l,l_+}\sqrt{2k_t}A_{in,+} e^{-ib_0t}\nonumber\\
&& + \delta_{l,l_-}\sqrt{2k_t}A_{in,-}e^{ib_0t} + ig_0\sum\limits_{m,n,p}\delta_{m-n,l-p} A_m A_p A^*_n.\ \ \ \ \ \
\end{eqnarray}
In the above, we have assumed that the loss rates are independent of mode number and used the more concise notation $2k = \Delta\omega_{tot}$, $2k_t = \Delta\omega_{ext}$, $2k_i = \Delta\omega_{int}$, and $A_{in,\pm} = \sqrt{\frac{P_{\pm}}{\hbar\omega_{\pm}}}$.

Let us now demonstrate the direct equivalence between the mode equations and the corresponding LLE presented in section \ref{LLEs}. Define $A(\theta,t) = \sum\limits_{l}A_l(t)e^{-il\theta}$, where $\theta$ is the angle that goes around the ring cavity and thus parameterises the position on the circumference. Substituting $\dot{A}_l$ into the equation for $\dot{A}$, we get
\begin{eqnarray}
\label{firstAeqn}
&&\dot{A} = -k A + i\sigma_0 A - i \sum\limits_{n\geq2} \frac{\zeta_n}{n!} i^n \frac{\partial^nA}{\partial\theta^n} + ig_0|A|^2A \nonumber\\
&& + \sqrt{2k_t}A_{in,+} e^{-ib_0t}e^{-il_+\theta} + \sqrt{2k_t}A_{in,-}e^{ib_0t}e^{-il_-\theta}.\ \ \ \
\end{eqnarray}
Note that the factor of $e^{-il_{\pm}\theta}$ in the driving terms arises from the Fourier transform. In addition, without loss of generality, assume that $l_-=-l_+$. Further, define $\tau=\theta/\zeta_1$, $\tau\in[0,t_R]$, and recall the approximate relation \cite{fromMiro}:
\begin{equation}
\zeta_k = - \zeta_1^k \frac{L \beta_k}{t_R}.
\end{equation}
We replace $\theta$ by $\tau$ and $\zeta_n$ by $\beta_n$ in (\ref{firstAeqn}), and get
\begin{eqnarray}
\label{secondAeqn}
&& \frac{\partial A}{\partial t} = -k A - i\delta_0/t_R A + i \frac{L}{t_R} \sum\limits_{n\geq2} \frac{\beta_n}{n!} i^n \frac{\partial^nA}{\partial\tau^n} \nonumber\\
&& + \sqrt{2k_t}A_{in,+} e^{ibt/t_R}e^{-i\Omega_p\tau} + \sqrt{2k_t}A_{in,-}e^{-ibt/t_R}e^{i\Omega_p\tau}\nonumber\\
&& + ig_0|A|^2A,
\end{eqnarray}
where we have also expressed the exponents through $\delta_0$ and $b$ (instead of $\sigma_0$ and $b_0$), and defined $\Omega_p=\zeta_1 l_+$. In order to complete the transformation, we identify parameters from the LLE of section \ref{LLEs} with their equivalents from the mode equations of this section:
\begin{eqnarray}
\label{ParamMap}
A &=& \sqrt{\frac{t_R}{\hbar \omega_0}}E,\nonumber\\
A_{in,\pm} &=& \frac{1}{\sqrt{\hbar \omega_0}}E_{in,\pm},\nonumber\\
t_Rk &=& \alpha, \nonumber\\
2k_t t_R &=& \tilde{\theta},\nonumber\\
g_0t_R^2/\hbar\omega_0 &=& \gamma L.
\end{eqnarray}
In order to rewrite our Eq.~(\ref{secondAeqn}) in the normalisation used in section \ref{LLEs}, we multiply (\ref{secondAeqn}) through by $\sqrt{\hbar \omega_0/t_R}$ and replace $g_0|A|^2$ by $\frac{\gamma L}{t_R}|E|^2$, which leads to
\begin{eqnarray}
&& \frac{\partial E}{\partial t} = -\frac{\alpha}{t_R} E - i\frac{\delta_0}{t_R} E + i \frac{L}{t_R} \sum\limits_{n\geq2} \frac{\beta_n}{n!} i^n \frac{\partial^n E}{\partial\tau^n} + i\frac{\gamma L}{t_R}|E|^2E\nonumber\\
&& + \frac{\sqrt{\tilde{\theta}}}{t_R} E_{in,+} e^{ibt/t_R}e^{-i\Omega_p\tau} + \frac{\sqrt{\tilde{\theta}}}{t_R} E_{in,-}e^{-ibt/t_R}e^{i\Omega_p\tau}.\ \ \ \ \ \ \ \ \
\end{eqnarray}
This is of course identical to Eq.~(\ref{biLLE}).

To conclude this section, it is useful to cast the mode equations in the same dimensionless form as our LLE (\ref{dimless_LLE}). Starting from (\ref{dimfulMode}), we use (\ref{ParamMap}) to first renormalise the variables from $A_l$ to $E_l$ (defined analogously to the relation between $A$ and in $E$ in (\ref{ParamMap})), and then take out units using (\ref{DimlessVars}), which leads to
\begin{eqnarray}
\label{dimlessMode}
\frac{\partial \tilde{E}_l}{\partial \tilde{t}} &=& -\tilde{E}_l - i\Delta_0 \tilde{E}_l + i \sum\limits_{n\geq2} \left(\frac{2\pi\tau_s}{t_R}\right)^n \frac{d_n}{n!} l^n \tilde{E}_l\nonumber\\
&+& \delta_{l,l_+}S_+ e^{ia\tilde{t}} + \delta_{l,l_-}S_-e^{-ia\tilde{t}}\nonumber\\
&+& i\sum\limits_{m,n,p}\delta_{m-n,l-p} \tilde{E}_m \tilde{E}_p \tilde{E}^*_n.
\end{eqnarray}
In practice, we solve the mode equations using the SSFM \cite{Hansson} \footnote{Note that the complex amplitudes obtained from the SSFM need to be divided by the total number of modes in order to obtain $\tilde{E}_l$ which satisfy the mode equations as stated.}, which is exactly equivalent to solving the corresponding LLE (for which we use the same method).
\subsection{Quantised model}
\label{Quantum}
\subsubsection{Linearised quantum fluctuations}
\label{LinQuantFlucts}
Having derived the mode equations for our bichromatically pumped system, we can now proceed to quantise them, analogously to what is done in \cite{Chembo}. We assume that the resonator is coupled to one fibre only, which both delivers the pump beams and carries away the signal coupled out of the cavity for measurement in the output beam (this configuration is referred to as ``add-through'' in \cite{Chembo}).

The classical mode equations are quantised by replacing the mode amplitudes $A_l$ by the lowering operators $\hat{a}_l$, obeying bosonic multi-particle statistics:
\begin{equation}
[\hat{a}_l,\hat{a}_p^{\dagger}]=\delta_{l,p},\ [\hat{a}_l,\hat{a}_p]=[\hat{a}_l^{\dagger},\hat{a}_p^{\dagger}]=0.
\end{equation}
Quantum fluctuations enter the system through the ports where the system interfaces with the environment. These are also the channels through which the system loses energy to the surroundings. In the add-through configuration, there are two such ports: ``intrinsic'' (labelled `i' below), accounting for irreversible dissipation, and the ``through-port'' (labelled `t' below), describing coupling of light in and out of the system via the fibre. Photons can be lost from the intracavity field to these two channels, which is accounted for in the classical model. However, at these ports, the system couples to the environment (assumed to be a vacuum), and quantum-mechanically, the vacuum fluctuations of the environment affect the intracavity mode fields by coupling directly to them.

In order to model the vacuum fluctuations of the environment, we need to introduce a set of bosonic operators to describe them. Let $\hat{V}_{s,l}(t)$ be a lowering operator, where the label `s' stands for either `i' or `t' to indicate the port, the label `l' indicates which cavity mode this operator couples to (in essence, this translates to the frequency of the field it describes), and the argument $t$ shows that we are considering the operator as time-dependant (i.e.~working in the Heisenberg picture). We can therefore define the Fourier transform of the vacuum operators and their adjoints as (see discussion of Eqs.~(\ref{FT1})-(\ref{FT2}), which is relevant here, too)
\begin{eqnarray}
\label{FT1b}
\hat{V}_{s,l}(\Omega) &=& \int dt\ \hat{V}_{s,l}(t) e^{-i\Omega t},\\
\hat{V}_{s,l}^{\dagger}(\Omega) &=& \int dt\ \hat{V}_{s,l}^{\dagger}(t) e^{i\Omega t}.
\label{FT2b}
\end{eqnarray}
The vacuum fluctuation operators have zero means and the following two-photon correlations, where expectation values are with respect to the vacuum state the environment is assumed to be in:
\begin{eqnarray}
&& \left[\hat{V}_{s,l}(t),\hat{V}_{s',l'}^{\dagger}(t')\right] = \left\langle \hat{V}_{s,l}(t) \hat{V}^{\dagger}_{s',l'}(t') \right\rangle = \nonumber\\
&& = \delta_{s,s'}\delta_{l,l'}\delta(t-t'),\nonumber\\
&& \left[ \hat{V}_{s,l}(t), \hat{V}_{s',l'}(t') \right] = \left[ \hat{V}^{\dagger}_{s,l}(t), \hat{V}^{\dagger}_{s',l'}(t') \right] = \nonumber\\
&& = \left\langle \hat{V}_{s,l}^{\dagger}(t) \hat{V}_{s',l'}(t') \right\rangle = \left\langle \hat{V}_{s,l}(t) \hat{V}_{s',l'}(t') \right\rangle = \nonumber\\
&& = \left\langle \hat{V}^{\dagger}_{s,l}(t) \hat{V}^{\dagger}_{s',l'}(t') \right\rangle = 0,\nonumber\\
&& \left[\hat{V}_{s,l}(\Omega),\hat{V}_{s',l'}^{\dagger}(\Omega')\right] = \left\langle \hat{V}_{s,l}(\Omega) \hat{V}^{\dagger}_{s',l'}(\Omega') \right\rangle = \nonumber\\
&& = \delta_{s,s'}\delta_{l,l'}\delta(\Omega-\Omega'),\nonumber\\
&& \left[ \hat{V}_{s,l}(\Omega), \hat{V}_{s',l'}(\Omega') \right] = \left[ \hat{V}^{\dagger}_{s,l}(\Omega), \hat{V}^{\dagger}_{s',l'}(\Omega') \right] = \nonumber\\
&& = \left\langle \hat{V}_{s,l}^{\dagger}(\Omega) \hat{V}_{s',l'}(\Omega') \right\rangle = \left\langle \hat{V}_{s,l}(\Omega) \hat{V}_{s',l'}(\Omega') \right\rangle = \nonumber\\
&& = \left\langle \hat{V}^{\dagger}_{s,l}(\Omega) \hat{V}^{\dagger}_{s',l'}(\Omega') \right\rangle = 0.
\label{Vproperties}
\end{eqnarray}
As usual, $\delta(t)$ is the Dirac delta function. The quantised mode equations (in full units) read
\begin{eqnarray}
&& \dot{\hat{a}}_l = -k \hat{a}_l - i\frac{\delta_0}{t_R} \hat{a}_l - i \sum\limits_{n\geq2} \frac{\zeta_n}{n!} l^n \hat{a}_l\nonumber\\
&& + \delta_{l,l_+}\sqrt{2k_t}A_{in,+} e^{ibt/t_R} + \delta_{l,l_-}\sqrt{2k_t}A_{in,-}e^{-ibt/t_R}\nonumber\\
&& + ig_0\sum\limits_{m,n,p}\delta_{m-n,l-p} \hat{a}^{\dagger}_n\hat{a}_m \hat{a}_p + \sqrt{2k_t}\hat{V}_{t,l} + \sqrt{2k_i}\hat{V}_{i,l}.\ \ \ \
\end{eqnarray}
We now linearise these equations. That is, we write $\hat{a}_l=A_l + \delta\hat{a}_l$, where the $\delta\hat{a}_l$'s are bosonic lowering operators, representing small quantum fluctuations with zero means on top of the classical means $A_l$. They obey the same commutation relations as the $\hat{a}_l$'s.

The differential equations for the classical means (i.e.~all the zeroth order terms in the fluctuations) are precisely the classical equations we derived in the previous section. The first order equations for the fluctuations are
\begin{eqnarray}
&& \delta\dot{\hat{a}}_l = -k \delta\hat{a}_l - i\frac{\delta_0}{t_R} \delta\hat{a}_l - i \sum\limits_{n\geq2} \frac{\zeta_n}{n!} l^n \delta\hat{a}_l\nonumber\\
&& + ig_0\sum\limits_{m,n,p} \delta_{m-n,l-p} \left( \delta\hat{a}^{\dagger}_n A_m A_p + A^*_n \delta\hat{a}_m A_p + A^*_n A_m \delta\hat{a}_p \right)\nonumber\\
&& + \sqrt{2k_t}\hat{V}_{t,l} + \sqrt{2k_i}\hat{V}_{i,l}.
\end{eqnarray}

It is useful to briefly also write down this equation in dimensionless units so that one can easily adapt the rest of the analysis (which we will do in full units) to a version that can be meaningfully coded (all our numerical simulations deal with dimensionless quantities). Clearly, the fluctuations $\delta\hat{a}_l$ scale the same way as the means $A_l$. However, the scaling of the vacuum fluctuation operators needs to be defined. Let $\hat{\tilde{V}}(\tilde{t})$ be the dimensionless operator of the dimensionless slow time, given by $\hat{\tilde{V}}(\tilde{t}) = \sqrt{t_R/\alpha}\hat{V}(t)$. The dimensionless equations for the quantum fluctuations are
\begin{eqnarray}
&& \delta\dot{\hat{\tilde{E}}}_l = - \delta\hat{\tilde{E}}_l - i\Delta_0 \delta\hat{\tilde{E}}_l + i \sum\limits_{n\geq2} \left(\frac{2\pi\tau_s}{t_R}\right)^n \frac{d_n}{n!} l^n \delta\hat{\tilde{E}}_l\nonumber\\
+ && i\sum\limits_{m,n,p} \delta_{m-n,l-p} \left( \delta\hat{\tilde{E}}^{\dagger}_n \tilde{E}_m \tilde{E}_p + \tilde{E}^*_n \delta\hat{\tilde{E}}_m \tilde{E}_p + \tilde{E}^*_n \tilde{E}_m \delta\hat{\tilde{E}}_p \right)\nonumber\\
+ && \sqrt{\frac{2k_t}{k}\frac{\gamma L}{\alpha} \frac{\hbar \omega_0}{t_R}} \hat{\tilde{V}}_{t,l} + \sqrt{\frac{2k_i}{k}\frac{\gamma L}{\alpha} \frac{\hbar \omega_0}{t_R}} \hat{\tilde{V}}_{i,l},
\label{dimlessflucts}
\end{eqnarray}
where $\delta\hat{\tilde{E}}_l$ are the quantum-mechanical fluctuations on top of the means $\tilde{E}_l$. The expectation values and commutators of the dimensionless $\hat{\tilde{V}}(\tilde{t})$'s are the same as the dimensionful $\hat{V}(t)$'s, except that all Dirac delta functions are of dimensionless time (and frequency, if in Fourier space), such that the operators and the arguments are unitless.

Note that while numerical simulations are performed in dimensionless units consistent with $\delta\hat{\tilde{E}}_l$, eventually we are interested in the fluctuations in the \textit{number amplitudes}, $\delta\hat{a}_l$, which are also dimensionless but scale differently. It is therefore necessary to convert between the two via $\delta\hat{a}_l = \sqrt{\frac{t_R}{\hbar\omega_0}\frac{\alpha}{\gamma L}} \delta\hat{\tilde{E}}_l$, so that the dimensionless scaling factors in front of the vacuum operators in (\ref{dimlessflucts}) cancel out and do not influence the final results at all. However, the ratio $k_t/k$ remains as an additional meaningful dimensionless parameter which is necessary to fully describe the quantum fluctuations but which does not feature in the classical equations. The case of critical coupling corresponds to $k_t/k=0.5$. Note that because $k=\alpha/t_R$ [see Eqs.~(\ref{ParamMap})] and fast time is scaled by $t_R/\alpha$ [see Eqs.~(\ref{DimlessVars})], the dimensionless version of $k$ is equal to one in our units.

Returning to the full-units analysis, the fluctuation equations can be conveniently written as
\begin{equation}
\delta\dot{\hat{a}}_l = \sum\limits_p R_{l,p} \delta \hat{a}_p + \sum\limits_p S_{l,p} \delta\hat{a}_p^{\dagger} + \sum\limits_s \sqrt{2k_s} \hat{V}_{s,l},
\end{equation}
where the $R$ and $S$ matrix elements are given by
\begin{eqnarray}
R_{l,p} &=& \delta_{l,p} \left(-k - i\delta_0/t_R - i \sum\limits_{n\geq2} \frac{\zeta_n}{n!} l^n\right)\nonumber\\
&+& 2ig_0\sum\limits_{m,n} \delta_{m-n,l-p} A^*_nA_m,\\
S_{l,p} &=& ig_0\sum\limits_{m,n} \delta_{m+n,l+p} A_nA_m.
\end{eqnarray}
This set of linear differential equations for the fluctuation operators can be solved by a Fourier transform as long as the coefficient matrices $R$ and $S$ are independent of time (note that one has to solve the classical equations for the state of interest first). Let the under-tilde symbol denote a vector of all indices running over mode number $l$. We construct block vectors and matrices which are twice the size of the system:
\begin{equation}
\left(
\begin{array}{c}
\delta \undertilde{\dot{\hat{a}}}\\
\delta \undertilde{\dot{\hat{a}}}^{\dagger}
\end{array}
\right)
=
\left(
\begin{array}{cc}
R & S\\
S^* & R^*
\end{array}
\right)
\left(
\begin{array}{c}
\delta \undertilde{\hat{a}}\\
\delta \undertilde{\hat{a}}^{\dagger}
\end{array}
\right)
+\sum\limits_s \sqrt{2k_s}
\left(
\begin{array}{c}
\undertilde{\hat{V}}_s\\
\undertilde{\hat{V}}^{\dagger}_s
\end{array}
\right).
\end{equation}
In the above equation, the fluctuation operators -- both of the cavity modes and of the vacuum -- are functions of time. We will now take the (forward) Fourier transform of this equation, but due to the already cumbersome notation, we will make clear that the operators are Fourier transforms by explicitly writing down their frequency dependence. For example, $\hat{V}$ is an operator in the time domain, but $\hat{V}(\Omega)$ is its Fourier transform. Observe that in order for the raising and lowering operators in frequency space to be adjoints, the lowering operator is defined as the forward Fourier transform, while the raising operator is defined as the inverse Fourier transform, according to
\begin{eqnarray}
\label{FT1}
\delta \hat{a}_l(\Omega) &=& \int dt\ \delta\hat{a}_l(t) e^{-i\Omega t},\\
\delta \hat{a}^{\dagger}_l(\Omega) &=& \int dt\ \delta\hat{a}^{\dagger}_l(t) e^{i\Omega t}.
\label{FT2}
\end{eqnarray}
Thus, taking the Fourier transform of the differential equations and solving the resulting algebraic equations gives
\begin{widetext}
\begin{equation}
\label{FTsoln}
\left(
\begin{array}{c}
\delta \undertilde{\hat{a}}(\Omega)\\
\delta \undertilde{\hat{a}}^{\dagger}(-\Omega)
\end{array}
\right)
=
-\left[\left(
\begin{array}{cc}
R & S\\
S^* & R^*
\end{array}
\right)
-i\Omega I\right]^{-1}
\sum\limits_s \sqrt{2k_s}
\left(
\begin{array}{c}
\undertilde{\hat{V}}_s(\Omega)\\
\undertilde{\hat{V}}^{\dagger}_s(-\Omega)
\end{array}
\right).
\end{equation}
\end{widetext}
Here, $I$ is the identity matrix, the same size as the composite matrix it is added to. Note that in order to arrive at Eq.~(\ref{FTsoln}), we have assumed that the classical mode amplitudes $A_l$ (on which matrices $R$ and $S$ depend) have reached steady-state and are constant in time. For future convenience, we will denote the matrix inverse appearing above as
\begin{equation}
J(\Omega) =
\left[\left(
\begin{array}{cc}
R & S\\
S^* & R^*
\end{array}
\right)
-i\Omega I\right]^{-1}.
\end{equation}
\subsubsection{Quadrature spectra}
\label{QuadSpecs}
Next, we will define two-mode quadrature operators which can be used to search for two-mode squeezing (once again, following \cite{Chembo}). The modes in question can be classically occupied, which makes it particularly useful for studying the properties of soliton states. The final quantity we will compute is the quadrature spectrum (defined below), which is none other than the variance of the field quadrature it corresponds to. As such, the quadrature spectrum is a natural quantity to inspect if one is interested in squeezing.

Thus, define for $l=1,...,K$
\begin{equation}
\label{q_eqn}
\delta \hat{q}_{\phi,l}(t) = \frac{1}{\sqrt{2}}(\delta \hat{a}_l(t) - \delta \hat{a}_{-l}(t))e^{-i\phi} + \mbox{H.c.},
\end{equation}
where ``H.c.'' stands for the Hermitian conjugate. Let $\delta\undertilde{\hat{q}}_{\phi}(t)$ be a vector of length $K$ with entries $\delta \hat{q}_{\phi,l}(t)$. When $\phi=0$, (\ref{q_eqn}) becomes the ``amplitude quadrature'', and when $\phi=\pi/2$, it reduces to the ``phase quadrature''. The quadratures at these two angles are particularly interesting, and their linear combination can give any other mixed-angle quadrature according to
\begin{equation}
\delta\undertilde{\hat{q}}_{\phi}(t) = \delta\undertilde{\hat{q}}_{0}(t)\cos(\phi) + \delta\undertilde{\hat{q}}_{\pi/2}(t)\sin(\phi).
\end{equation}
These two key quadrature fields are explicitly given by
\begin{eqnarray}
\delta \hat{q}_{0,l}(t) = \frac{1}{\sqrt{2}}(\delta \hat{a}_l(t) - \delta \hat{a}_{-l}(t)) + \mbox{H.c.},\\
\delta \hat{q}_{\pi/2,l}(t) = -\frac{i}{\sqrt{2}}(\delta \hat{a}_l(t) - \delta \hat{a}_{-l}(t)) + \mbox{H.c.}.
\end{eqnarray}
We would like to find the solutions in Fourier space for the above two operators. The differential equations for their temporal evolution can be trivially obtained by differentiating the RHSs. Taking the forward Fourier transform of both sides and recalling Eqs.~(\ref{FT1})-(\ref{FT2}), we have
\begin{eqnarray}
\delta \hat{q}_{0,l}(\Omega) &=& \frac{1}{\sqrt{2}}(\delta \hat{a}_l(\Omega) - \delta \hat{a}_{-l}(\Omega))\nonumber\\
&+& \frac{1}{\sqrt{2}}(\delta \hat{a}^{\dagger}_l(-\Omega) - \delta \hat{a}^{\dagger}_{-l}(-\Omega)),\nonumber\\
\delta \hat{q}_{\pi/2,l}(\Omega) &=& -\frac{i}{\sqrt{2}}(\delta \hat{a}_l(\Omega) - \delta \hat{a}_{-l}(\Omega))\nonumber\\
&+& \frac{i}{\sqrt{2}}(\delta \hat{a}^{\dagger}_l(-\Omega) - \delta \hat{a}^{\dagger}_{-l}(-\Omega)).
\end{eqnarray}
Since we already know the solutions for the $\delta \hat{a}_l(\Omega)$ and $\delta \hat{a}^{\dagger}_l(-\Omega)$ operators, the frequency-space quadrature fluctuation operators can be readily constructed \footnote{Note that the quadrature operators in frequency space are not Hermitian, as they are in the time domain.}.

The output quadrature field operators are
\begin{eqnarray}
&& \delta \hat{Q}_{out,0,l}(t) = \sqrt{2k_t}\delta \hat{q}_{0,l}(t) - \frac{1}{\sqrt{2}}\left[ \hat{V}_{t,l}(t) - \hat{V}_{t,-l}(t) \right]\nonumber\\
&& - \frac{1}{\sqrt{2}}\left[ \hat{V}_{t,l}^{\dagger}(t) - \hat{V}_{t,-l}^{\dagger}(t) \right],\\
&& \delta \hat{Q}_{out,\pi/2,l}(t) = \sqrt{2k_t}\delta \hat{q}_{\pi/2,l}(t) + \frac{i}{\sqrt{2}}\left[ \hat{V}_{t,l}(t) - \hat{V}_{t,-l}(t) \right]\nonumber\\
&& - \frac{i}{\sqrt{2}} \left[ \hat{V}_{t,l}^{\dagger}(t) - \hat{V}_{t,-l}^{\dagger}(t) \right].
\end{eqnarray}
Taking the Fourier transform of this expression, and recalling Eqs.~(\ref{FT1b})-(\ref{FT2b}), we get
\begin{eqnarray}
&& \delta \hat{Q}_{out,0,l}(\Omega) = \sqrt{2k_t}\delta \hat{q}_{0,l}(\Omega) - \frac{1}{\sqrt{2}}\left[ \hat{V}_{t,l}(\Omega) - \hat{V}_{t,-l}(\Omega)\right]\nonumber\\
&& - \frac{1}{\sqrt{2}}\left[ \hat{V}_{t,l}^{\dagger}(-\Omega) - \hat{V}_{t,-l}^{\dagger}(-\Omega) \right],\\
&& \delta \hat{Q}_{out,\pi/2,l}(\Omega) = \sqrt{2k_t}\delta \hat{q}_{\pi/2,l}(\Omega) + \frac{i}{\sqrt{2}}\left[ \hat{V}_{t,l}(\Omega) - \hat{V}_{t,-l}(\Omega) \right]\nonumber\\
&& - \frac{i}{\sqrt{2}}\left[ \hat{V}_{t,l}^{\dagger}(-\Omega) - \hat{V}_{t,-l}^{\dagger}(-\Omega) \right].
\end{eqnarray}
We wish to compute
\begin{equation}
\label{QuadSpec}
(X,Y)_l = \int \left\langle\delta \hat{Q}_{out,X,l}(\Omega) \delta \hat{Q}_{out,Y,l}(\Omega')\right\rangle d\Omega',
\end{equation}
where $l\in[1,K]$, and $X,Y\in\{0,\pi/2\}$. In order to compute this kind of spectrum at an arbitrary angle $\phi$, one must take the linear combination
\begin{eqnarray}
S_{\phi,l} &=& (0,0)_l\cos^2(\phi) + (\pi/2,\pi/2)_l\sin^2(\phi)\nonumber\\
&+& [(0,\pi/2)_l + (\pi/2,0)_l]\cos(\phi)\sin(\phi).
\label{SqzngSpec}
\end{eqnarray}

Let $x,y\in \{1,-i\}$ correspond to phase $X,Y\in\{0,\pi/2\}$. That is, if $X=0$, then $x=1$, and if $Y=\pi/2$, then $y=-i$ (i.e.~we have a state label and a parameter \textit{pair} that go together). After some algebra, we get the final result
\begin{widetext}
\begin{eqnarray}
(X,Y)_l &=& 2kk_t\left\{ \sum\limits_{i=1}^{2K+1} \left[-xJ_{K+1+l,i}(+\Omega) + xJ_{K+1-l,i}(+\Omega)\right]\times \left[-yJ_{K+1+l,2K+1+i}(-\Omega) + yJ_{K+1-l,2K+1+i}(-\Omega)\right]\right. \nonumber\\
&+& \sum\limits_{i=1}^{2K+1} \left[-xJ_{K+1+l,i}(+\Omega) + xJ_{K+1-l,i}(+\Omega)\right]\times \left[-y^*J_{3K+2+l,2K+1+i}(-\Omega) + y^*J_{3K+2-l,2K+1+i}(-\Omega)\right] \nonumber\\
&+& \sum\limits_{i=1}^{2K+1} \left[-x^*J_{3K+2+l,i}(+\Omega) + x^*J_{3K+2-l,i}(+\Omega)\right] \left[-yJ_{K+1+l,2K+1+i}(-\Omega) + yJ_{K+1-l,2K+1+i}(-\Omega)\right] \nonumber\\
&+& \left. \sum\limits_{i=1}^{2K+1} \left[-x^*J_{3K+2+l,i}(+\Omega) + x^*J_{3K+2-l,i}(+\Omega)\right]\times \left[-y^*J_{3K+2+l,2K+1+i}(-\Omega) + y^*J_{3K+2-l,2K+1+i}(-\Omega)\right] \right\} \nonumber\\
&+& k_t x y \left[ J_{K+1+l,3K+2+l}(-\Omega) - J_{K+1-l,3K+2+l}(-\Omega) - J_{K+1+l,3K+2-l}(-\Omega) + J_{K+1-l,3K+2-l}(-\Omega) \right] \nonumber\\
&+& k_t x y^{*} \left[ J_{3K+2+l,3K+2+l}(-\Omega) - J_{3K+2-l,3K+2+l}(-\Omega) - J_{3K+2+l,3K+2-l}(-\Omega) + J_{3K+2-l,3K+2-l}(-\Omega) \right] \nonumber\\
&+& k_t x y^{*} \left[ J_{K+1+l,K+1+l}(+\Omega) - J_{K+1+l,K+1-l}(+\Omega) - J_{K+1-l,K+1+l}(+\Omega) + J_{K+1-l,K+1-l}(+\Omega)\right] \nonumber\\
&+& k_t x^{*} y^{*} \left[ J_{3K+2+l,K+1+l}(+\Omega) - J_{3K+2+l,K+1-l}(+\Omega) - J_{3K+2-l,K+1+l}(+\Omega) + J_{3K+2-l,K+1-l}(+\Omega) \right] \nonumber\\
&+& x y^{*}.
\end{eqnarray}
\end{widetext}

An important remark is in order. The classical amplitudes $A_{\pm l}$ are in general complex and have phases. This has the effect of shifting the phase of the computed quadratures, such that the amplitude quadrature is no longer found at $\phi=0$ and the phase quadrature is no longer at $\phi=\pi/2$ (however, the two are still $\pi/2$ apart). Instead, the amplitude quadrature can be found at the average value of the phases of $A_l$ and $A_{-l}$, as discussed in \cite{Chembo}, and the phase quadrature angle would then differ by $\pi/2$ from this value. In general, $S_{\phi,l}$ has a divergence at $\Omega=0$, and only at the special angle which correctly yields the amplitude quadrature is the divergence completely absent. Note that $\phi_{min}$ defined in section \emph{Two-mode squeezing} of the main text does not in general coincide with this angle.

The results in Fig.~4 of the main manuscript were obtained by applying the mathematical framework above. The optimal angle $\phi_{min}$ is identified as follows. For every mixing angle $\phi$, we find the smallest value attained by the quadrature spectrum of mode pair $l$, $S_{\phi,l}$ over all $\Omega$. We then plot the minimal value of $S_{\phi,l}$ as a function of $\phi$. The angle at which the minimum (per spectrum) is the smallest (as a function of $\phi$) is denoted as $\phi_{min}$.

The curves shown in Fig.~4(a) of the main text are computed from (\ref{SqzngSpec}), setting $l=1$ and $\phi=\phi_{min}$, while the results in Fig.~4(b) extract the value of the squeezing spectrum at the lowest point (as a function of $\Omega$) for a range of mode numbers $l$, each spectrum plotted at its own $\phi_{min}$.
\subsection{Timing jitter spectrum}
\label{Jitter_Method}
In reality, there are many sources of noise in the system, both classical and quantum mechanical. In this section, we will completely ignore all classical noise, as well as any quantum fluctuations introduced to the system from the laser producing the beam(s) driving the resonator. We will only consider noise resulting from the interaction of the system with the quantised vacuum via the ``through-port-'' and ``intrinsic-'' loss channels, as studied thus far in the above. Our goal is to obtain the fundamental timing-jitter spectrum of PDCSs, and compare it to that of ordinary CSs, which has been studied previously \cite{jitter, quantum_solitons4}. First, we will take an analytical approach analogous to that presented in \cite{jitter} and derive an explicit formula for the spectrum, and then we will provide a description of numerical simulations modelling the system and explain how the spectrum can be extracted from these. In this regard, the article \cite{NewHope} contains extremely useful guidelines, some of which we will also adopt.
\subsubsection{Analytical derivation}
\label{Jitter_Anal}
One way to access the timing jitter spectrum is by following the method of \cite{jitter}, where the authors have carried out the derivation for ordinary CSs. Here we will give an outline of the calculation for PDCSs and simply restate the final result for CSs to enable direct comprison. We will work in the same dimensionless units we have used in the rest of the Supplementary Information, which will facilitate comparison to simulations described later on. We will use the simplified model of Eq.~(\ref{simple_LLE}) to describe the PDCS, as it accurately captures the soliton itself \cite{PDCS}, while information about the driving fields and interference between them which is included in the full model (\ref{dimless_LLE}) and is absent from the simple model (\ref{simple_LLE}) is not needed to correctly predict the timing jitter. We validate this intuitive understanding explicitly, using numerical simulations, in the main text, section \emph{Fundamental timing jitter}.

The derivation can be divided into two parts: first, one employs the ``Lagrangian method'' to derive ordinary differential equations (ODEs) for parameters entering an ansatz for the soliton, the steady state solution of which accurately captures the soliton solution of the governing LLE. Second, one adds quantum fluctuations to the model, and modifies the ODEs for the parameters to include these new drive terms. Linearised deviations from the steady state solution for the parameters driven by the quantum fluctuations can be solved for exactly, which consequently allows for the computation of the jitter spectrum.

Let us thus begin from the Lagrangian method. It proceeds as follows. Consider the simplified LLE describing PDCSs, Eq.~(\ref{simple_LLE}), and assume that only second order dispersion is included. The slow-time-derivative term, the dispersion, nonlinearity, and detuning terms are all ``conservative'' -- they describe processes that do not change the overall energy in the system. The loss and drive terms, on the other hand, are not conservative: they describe energy transfer between the environment and the system. If for the moment we only keep the conservative terms of the LLEs, then they can be captured by a Lagrangian density
\begin{eqnarray}
\mathcal{L} &=& \frac{i}{2}\left(\tilde{E}_S^*\frac{\partial \tilde{E}_S}{\partial \tilde{t}} - \tilde{E}_S\frac{\partial \tilde{E}_S^*}{\partial \tilde{t}} \right) + \frac{d_2}{2}\frac{\partial \tilde{E}_S^*}{\partial \tilde{\tau}} \frac{\partial \tilde{E}_S}{\partial \tilde{\tau}}\nonumber\\
&+& \frac{1}{2}|\tilde{E}_S|^4 - \Delta_{\mbox{eff}} |\tilde{E}_S|^2.
\end{eqnarray}
The Euler-Lagrange equation derived from this Lagrangian density is none other than the conservative part of the LLE:
\begin{eqnarray}
&& \frac{\partial \mathcal{L}}{\partial \tilde{E}_S^*} - \frac{\partial}{\partial \tilde{\tau}} \frac{\partial \mathcal{L}}{\partial(\partial \tilde{E}_S^*/\partial\tilde{\tau})} - \frac{\partial}{\partial \tilde{t}} \frac{\partial \mathcal{L}}{\partial(\partial \tilde{E}_S^*/\partial \tilde{t})} = \nonumber\\
\label{conservative_LLE}
&& i \frac{\partial \tilde{E}_S}{\partial \tilde{t}} + |\tilde{E}_S|^2\tilde{E}_S - \Delta_{\mbox{eff}} \tilde{E}_S - \frac{d_2}{2} \frac{\partial^2 \tilde{E}_S}{\partial \tilde{\tau}^2}=0.
\end{eqnarray}
The full LLE of interest -- including the non-conservative terms -- has the zero on the RHS replaced by $\mathcal{R}$, where $\mathcal{R} = -2\tilde{E}_+\tilde{E}_-\tilde{E}_S^* - i\tilde{E}_S$.

Next, one writes down an ansatz for the soliton solution of the LLE
\begin{equation}
\label{ansatz2}
\tilde{E}_S(\tilde{t},\tilde{\tau}) = A\ \mbox{sech}\left(\frac{\tilde{\tau}-P}{w}\right) e^{-i\omega_s(\tilde{\tau}-P)}e^{i\phi}.
\end{equation}
Here, $A$ is the amplitude, $P$ is the soliton position, $w$ is its width, $\omega_s$ is the central frequency and $\phi$ the phase. Substituting ansatz (\ref{ansatz2}) into the conservative LLE (\ref{conservative_LLE}), we obtain two conditions the parameters must satisfy:
\begin{eqnarray}
|A|^2 + \frac{d_2}{w^2}=0,\\
\Delta_{\mbox{eff}} + \frac{d_2}{2w^2}=0.
\end{eqnarray}
The first condition establishes a general relation between the width and height of the soliton. This can be understood as energy conservation for the soliton, since the (fast-time) integral over the intensity profile is proportional to the enegry contained in the soliton. Note that this relation is usually maintained as a function of slow time even when the non-conservative terms are added to the LLE. The second condition establishes a concrete value for the soliton width in terms of the detuning, and this equation is only valid for the conservative LLE, so it is of no further interest to us. In addition, in order for (\ref{ansatz2}) to satisfy (\ref{conservative_LLE}) we need $P=0$, $\omega_s=0$, and $\phi$ can be arbitrary. We will see that these three conditions will remain true even when non-conservative terms are accounted for.

The next step is precisely to add the non-conservative terms into the LLE and account for their effect on the parameters by deriving ODEs that govern the evolution of the soliton parameters as a function of slow time, eventually converging to their steady-state values (that agree with direct solutions of the LLE by conventional means). For our system, in order to obtain meaningful (and correct) results, one should not allow all the parameters to vary simultaneously: either we fix $P=0$, $\omega_s=0$, $w=\sqrt{-d_2}/A$ and allows $A$ and $\phi$ to vary with slow time (once $\mathcal{R}$ is introduced), \textbf{or} one assumes $A$, $w$ and $\phi$ are fixed at their steady-state values (corresponding to the solution of the full, non-conservative LLE) and derives ODEs for $\omega_s$ and $P$ instead.

Since our ultimate goal is to study the dynamics of $P$ (with quantum fluctuations introduced, as well), we need an ODE for $P$. Thus, in this derivation, we will proceed with the second option listed above. We assume that three parameters are fixed at their steady-state values: $A_{ss}$, $w_{ss}$, $\phi_{ss}$. Their exact values can be obtained either by reading off the soliton properties from a numerical solution to the full LLE, or directly from the analytical solution (\ref{analPDCS}). In contrast, we assume $P$ and $\omega_s$ can change with slow time.

In order to proceed, we must evaluate $\mathcal{L}$ for the ansatz (\ref{ansatz2}), treating three of the parameters as constants ($A_{ss}$, $w_{ss}$, $\phi_{ss}$) and two as functions of slow time ($P$ and $\omega_s$). Integrating $\mathcal{L}$ over $\tilde{\tau}$ yields the Lagrangian
\begin{eqnarray}
L &=& \frac{d_2A_{ss}^2}{3w_{ss}} + d_2\omega_s^2A_{ss}w_{ss} + \frac{2A_{ss}^4w_{ss}}{3}\nonumber\\
&-& 2\Delta_{\mbox{eff}} A_{ss}^2w_{ss} - 2\omega_s A_{ss}^2w_{ss}\dot{P},
\end{eqnarray}
where an over-dot indicates a derivative with respect to $\tilde{t}$. The ODEs governing the dynamics of the soliton parameters can be obtained from
\begin{equation}
\label{ODEs}
\frac{\partial L}{\partial r} -\frac{d}{d\tilde{t}}\left(\frac{\partial L}{\partial \dot{r}}\right) = \int d\tilde{\tau}\ \mathcal{R}\frac{\partial \tilde{E}_S^*}{\partial r} + \mathcal{R}^*\frac{\partial \tilde{E}_S}{\partial r},
\end{equation}
where $r$ is either $P$ or $\omega_s$. The left-hand side (LHS) of (\ref{ODEs}) only depends on $L$. In particular
\begin{eqnarray}
&& \frac{\partial L}{\partial P} -\frac{d}{d\tilde{t}}\left(\frac{\partial L}{\partial \dot{P}}\right) = 2A_{ss}^2w_{ss}\dot{\omega}_s,\\
&& \frac{\partial L}{\partial \omega_s} -\frac{d}{d\tilde{t}}\left(\frac{\partial L}{\partial \dot{\omega}_s}\right) = -2A_{ss}^2w_{ss}\dot{P} + 2\omega_s d_2A_{ss}^2w_{ss}.\ \ \ \
\end{eqnarray}
The RHS of (\ref{ODEs}), on the other hand, depends on $\mathcal{R}$. For PDCSs, we have
\begin{eqnarray}
&& \int d\tilde{\tau}\ \mathcal{R}\frac{\partial \tilde{E}_S^*}{\partial P} + \mathcal{R}^*\frac{\partial \tilde{E}_S}{\partial P} = -4A_{ss}^2w_{ss}\omega_s,\\
&& \int d\tilde{\tau}\ \mathcal{R}\frac{\partial \tilde{E}_S^*}{\partial \omega_s} + \mathcal{R}^*\frac{\partial \tilde{E}_S}{\partial \omega_s} = 4A_{ss}^2\pi w_{ss}^2 \nonumber\\
&& \times [-1+\omega_s \pi w_{ss}\mbox{coth}(\omega_s \pi w_{ss})]\mbox{cosech}(\omega_s \pi w_{ss}) \nonumber\\
&& \times [\mathcal{R}(\tilde{E}_+ \tilde{E}_-)\cos(2\phi_{ss}) + \mathcal{I}(\tilde{E}_+ \tilde{E}_-)\sin(2\phi_{ss})].
\end{eqnarray}
Constructing the ODEs by setting LHS equal to RHS leads to the steady-state solution $P=0,\ \omega_s=0$, which is of course correct and expected.

Having obtained the ODEs governing the classical dynamics of the soliton parameters, we can now add quantum fluctuations to the model in order to include their effect. As we have seen in Eq.~(\ref{dimlessflucts}), the dimensionless mode equations for $\frac{\partial \hat{\tilde{E}}_l}{\partial \tilde{t}}$ (note that $\hat{\tilde{E}}_l$ include the classical means as well as the quantum fluctuations) gain two new terms on the RHS
\begin{equation}
\label{freq_dom}
\sqrt{\frac{2k_t}{k}\frac{\gamma L}{\alpha} \frac{\hbar \omega_0}{t_R}} \hat{\tilde{V}}_{t,l}(\tilde{t}) + \sqrt{\frac{2k_i}{k}\frac{\gamma L}{\alpha} \frac{\hbar \omega_0}{t_R}} \hat{\tilde{V}}_{i,l}(\tilde{t}).
\end{equation}
The corresponding dimensionless quantised LLE -- the generalisation of Eq.~(\ref{simple_LLE}) -- also gains two new terms on the RHS:
\begin{equation}
\label{time_dom}
\sqrt{\frac{2k_t}{k}\frac{\gamma L}{\alpha} \frac{\hbar \omega_0}{t_R}} \hat{\tilde{V}}_{t}(\tilde{\tau},\tilde{t}) + \sqrt{\frac{2k_i}{k}\frac{\gamma L}{\alpha} \frac{\hbar \omega_0}{t_R}} \hat{\tilde{V}}_{i}(\tilde{\tau},\tilde{t}),
\end{equation}
with the connection between the vacuum operators in the fast-temporal and spectral domains being
\begin{equation}
\hat{\tilde{V}}_{s}(\tilde{\tau},\tilde{t}) = \sum\limits_{l} \hat{\tilde{V}}_{s,l}(\tilde{t}) e^{-2\pi i l \tilde{\tau}/\tilde{T}}.
\end{equation}
We have already mentioned that the properties of the dimentionless vacuum operators are the same as those in Eq.~(\ref{Vproperties}), except all the operators, slow time and frequency, are all replaced by their dimensionless versions. Likewise, the dimensionless vacuum operators in the temporal domain obey
\begin{eqnarray}
&& \left[\hat{\tilde{V}}_{s}(\tilde{\tau},\tilde{t}),\hat{\tilde{V}}_{s'}^{\dagger}(\tilde{\tau}',\tilde{t}')\right] = \left\langle \hat{\tilde{V}}_{s}(\tilde{\tau},\tilde{t}) \hat{\tilde{V}}^{\dagger}_{s'}(\tilde{\tau}',\tilde{t}') \right\rangle = \nonumber\\
&& = \delta_{s,s'}\delta\left(2\pi\frac{\tilde{\tau}-\tilde{\tau}'}{\tilde{T}}\right)\delta(\tilde{t}-\tilde{t}'),\nonumber\\
&& \left[ \hat{\tilde{V}}_{s}(\tilde{\tau},\tilde{t}), \hat{\tilde{V}}_{s'}(\tilde{\tau}',\tilde{t}') \right] = \left[ \hat{\tilde{V}}^{\dagger}_{s}(\tilde{\tau},\tilde{t}), \hat{\tilde{V}}^{\dagger}_{s'}(\tilde{\tau}',\tilde{t}') \right] = \nonumber\\
&& = \left\langle \hat{\tilde{V}}_{s}^{\dagger}(\tilde{\tau},\tilde{t}) \hat{\tilde{V}}_{s'}(\tilde{\tau}',\tilde{t}') \right\rangle = \left\langle \hat{\tilde{V}}_{s}(\tilde{\tau},\tilde{t}) \hat{\tilde{V}}_{s'}(\tilde{\tau}',\tilde{t}') \right\rangle = \nonumber\\
&& = \left\langle \hat{\tilde{V}}^{\dagger}_{s}(\tilde{\tau},\tilde{t}) \hat{\tilde{V}}^{\dagger}_{s'}(\tilde{\tau}',\tilde{t}') \right\rangle = 0,\nonumber\\
&& \left[\hat{\tilde{V}}_{s}(\tilde{\tau},\tilde{\Omega}),\hat{\tilde{V}}_{s'}^{\dagger}(\tilde{\tau}',\tilde{\Omega}')\right] = \left\langle \hat{\tilde{V}}_{s}(\tilde{\tau},\tilde{\Omega}) \hat{\tilde{V}}^{\dagger}_{s'}(\tilde{\tau}',\tilde{\Omega}') \right\rangle = \nonumber\\
&& = \delta_{s,s'}\delta\left(2\pi\frac{\tilde{\tau}-\tilde{\tau}'}{\tilde{T}} \right)\delta(\tilde{\Omega}-\tilde{\Omega}'),\nonumber\\
&& \left[ \hat{\tilde{V}}_{s}(\tilde{\tau},\tilde{\Omega}), \hat{\tilde{V}}_{s'}(\tilde{\tau}',\tilde{\Omega}') \right] = \left[ \hat{\tilde{V}}^{\dagger}_{s}(\tilde{\tau},\tilde{\Omega}), \hat{\tilde{V}}^{\dagger}_{s'}(\tilde{\tau}',\tilde{\Omega}') \right] = \nonumber\\
&& = \left\langle \hat{\tilde{V}}_{s}^{\dagger}(\tilde{\tau},\tilde{\Omega}) \hat{\tilde{V}}_{s'}(\tilde{\tau}',\tilde{\Omega}') \right\rangle = \left\langle \hat{\tilde{V}}_{s}(\tilde{\tau},\tilde{\Omega}) \hat{\tilde{V}}_{s'}(\tilde{\tau}',\tilde{\Omega}') \right\rangle = \nonumber\\
&& = \left\langle \hat{\tilde{V}}^{\dagger}_{s}(\tilde{\tau},\tilde{\Omega}) \hat{\tilde{V}}^{\dagger}_{s'}(\tilde{\tau}',\tilde{\Omega}') \right\rangle = 0.
\label{Vproperties2}
\end{eqnarray}
The fact that the LLE of interest now contains the vacuum fluctuation operators implies that $\mathcal{R}$ gains two new terms:
\begin{equation}
i\sqrt{\frac{2k_t}{k}\frac{\gamma L}{\alpha} \frac{\hbar \omega_0}{t_R}} \hat{\tilde{V}}_{t}(\tilde{\tau},\tilde{t}) + i\sqrt{\frac{2k_i}{k}\frac{\gamma L}{\alpha} \frac{\hbar \omega_0}{t_R}} \hat{\tilde{V}}_{i}(\tilde{\tau},\tilde{t}) \equiv \hat{\mathcal{R}}_V.
\end{equation}
These new terms give rise to new contributions to the RHS of Eq.~(\ref{ODEs}). Let
\begin{equation}
f_r(\tilde{t},\tilde{\tau}) = \hat{\mathcal{R}}_V\frac{\partial \tilde{E}_S^*}{\partial r} + \hat{\mathcal{R}}^{\dagger}_V\frac{\partial \tilde{E}_S}{\partial r},
\end{equation}
where $r=P$ or $\omega_s$, and
\begin{equation}
F_r(\tilde{t}) = \int d\tilde{\tau}\ f_r(\tilde{t},\tilde{\tau}).
\end{equation}
We must add $F_r(\tilde{t})$ to the RHS of (\ref{ODEs}) for parameter $r$, and set it equal to the LHS (which is unchanged), thus modifying the ODEs for the two variational parameters.

The $\hat{\mathcal{R}}_V$ terms are small (as they describe quantum fluctuations), so they will induce a small deviation $\delta r$ of the solution $r$ from its steady-state value $r_{ss}$. Thus, for each parameter $r$, we write $r = r_{ss} + \delta r$, where $r$ is either $P$ or $\omega_s$. We then isolate the terms that are linear in the fluctuations, driven by the noise terms $F_r$. Note that for PDCSs, while the $P$ equation is linear, the $\omega_s$ equation needs to be linearised by Taylor expanding non-linear functions. Define
\begin{eqnarray}
\mathcal{E} &=& 2A_{ss}^2w_{ss},\\
\label{lmn_l}
\zeta &=& d_2 - \frac{2}{3}\pi^2 w_{ss}^2 m,\\
\label{lmn_m}
m &=& \mathcal{R}(\tilde{E}_+ \tilde{E}_-)\cos(2\phi_{ss}) + \mathcal{I}(\tilde{E}_+ \tilde{E}_-)\sin(2\phi_{ss}).\ \ \ \ \
\end{eqnarray}
Then the ODEs for the fluctuations of PDCSs read
\begin{eqnarray}
\delta\dot{\omega}_s &=& -2\delta\omega_s + \frac{F_P(\tilde{t})}{\mathcal{E}},\\
\delta\dot{P} &=& \zeta\delta\omega_s - \frac{F_{\omega_s}(\tilde{t})}{\mathcal{E}}.
\end{eqnarray}
In order to solve these, we take the Fourier transform, turning differential equations in $\tilde{t}$ to algebraic equations in $\tilde{\Omega}$. Let $\delta r(\tilde{\Omega})$ and $F_r(\tilde{\Omega})$ be the Fourier transforms of $\delta r(\tilde{t})$ and $F_r(\tilde{t})$, respectively. Eliminating $\delta\omega_s(\tilde{\Omega})$, we get for PDCSs
\begin{equation}
\delta P(\tilde{\Omega}) = \frac{\zeta}{i\tilde{\Omega}\mathcal{E}}\frac{F_P(\tilde{\Omega})}{2+i\tilde{\Omega}} - \frac{F_{\omega_s}(\tilde{\Omega})}{i\tilde{\Omega}\mathcal{E}}.
\end{equation}
The timing jitter spectrum is defined as
\begin{equation}
\label{jitter_spec_def}
\int d\tilde{\Omega}'\ \left\langle \delta P^{\dagger}(\tilde{\Omega}') \delta P(\tilde{\Omega}) \right\rangle.
\end{equation}
In order to compute this, we need the result
\begin{equation}
\label{expectation}
\left\langle F_s^{\dagger}(\tilde{\Omega})F_r(\tilde{\Omega}')\right\rangle = \frac{\gamma L}{\alpha} \frac{\hbar \omega_0}{t_R} 2\delta(\tilde{\Omega}-\tilde{\Omega}')\int d\tilde{\tau}\ \frac{\partial \tilde{E}_S^*}{\partial s}(\tilde{\tau}) \frac{\partial \tilde{E}_S}{\partial r} (\tilde{\tau}),
\end{equation}
where $\tilde{E}_S$ is the ansatz (\ref{ansatz2}), and $s,r\in\{P,\omega_s\}$. Now, when we substitute the solutions for $\delta P(\tilde{\Omega})$ into (\ref{jitter_spec_def}), we get four terms of the form (\ref{expectation}). The two ``cross-terms'' where $s\neq r$ give rise to a term in the final spectrum which is asymmetric in $\tilde{\Omega}$, and is thus unphysical. It is usually dropped \cite{jitter} on these grounds, and as we shall do likewise, for brevity, we omit the details of its computetion. The remaining two terms for which $s=r$ require the following integrals
\begin{eqnarray}
\int d\tilde{\tau}\ \frac{\partial \tilde{E}_S^*}{\partial P}(\tilde{\tau}) \frac{\partial \tilde{E}_S}{\partial P} (\tilde{\tau}) &=& \frac{\mathcal{E}}{3w_{ss}^2} + \mathcal{E}\tilde{\Omega}^2,\\
\int d\tilde{\tau}\ \frac{\partial \tilde{E}_S^*}{\partial \omega_s}(\tilde{\tau}) \frac{\partial \tilde{E}_S}{\partial \omega_s} (\tilde{\tau}) &=& \frac{\pi^2w_{ss}^3A_{ss}^2}{6}.
\end{eqnarray}
The final expression for the timing jitter spectrum of PDCSs is
\begin{equation}
\label{CS_jitter_formula}
S_P = \frac{\gamma L \hbar \omega_0}{\alpha t_R} \frac{2}{3\mathcal{E}\tilde{\Omega}^2}\left(\frac{\zeta^2}{w_{ss}^2(4+\tilde{\Omega}^2)} + \frac{\pi^2 w_{ss}^2}{4}\right),
\end{equation}
to be compared with that of ordinary CSs \cite{jitter}:
\begin{equation}
\label{PDCS_jitter_formula}
S_P = \frac{\gamma L \hbar \omega_0}{\alpha t_R} \frac{2}{3\mathcal{E}\tilde{\Omega}^2}\left(\frac{d_2^2}{w_{ss}^2(4+\tilde{\Omega}^2)} + \frac{\pi^2 w_{ss}^2}{4}\right).
\end{equation}
We see that the spectrum is given by nearly the same expression, except that for PDCSs, $d_2$ is generalised to $\zeta$ (this is used in Eq.~(5) of the main text to compactly write down a single expression for both cases).

To complete the derivation, it is worth mentioning that in order to restore full units to $S_P$, one must multiply it by $t_R^2\tau_s/\alpha$, which follows from the fact that $\delta P(\tilde{t})$ needs a factor of $\sqrt{\tau_s t_R}$ to convert it to a dimensionful quantity. It may seem counter intuitive that $\delta P(t)$ scales differently from $\tau$ (which is in units of $\tau_s$), but this is because the jitter dynamics is driven by the quantum noise, and ultimately, it is the noise which determines the scaling of the soliton position deviation.

Finally, for convenience, we write down the jitter spectra in full dimensional units. For ordinary CSs it is
\begin{equation}
S_P = \frac{\hbar \omega_0 \alpha \tau_s^2}{3 A_{ss}^2 w_{ss} t_R \Omega^2} \left(\frac{4\tau_s^2}{w_{ss}^2(4+\Omega^2 t_R^2/\alpha^2)} + \frac{\pi^2 w_{ss}^2}{4\tau_s^2}\right),
\end{equation}
and for PDCSs, it is
\begin{equation}
S_P = \frac{\hbar \omega_0 \alpha \tau_s^2}{3 A_{ss}^2 w_{ss} t_R \Omega^2} \left(\frac{\zeta^2 \tau_s^2}{w_{ss}^2(4+\Omega^2 t_R^2/\alpha^2)} + \frac{\pi^2 w_{ss}^2}{4\tau_s^2}\right).
\end{equation}

In contrast to the notation in the rest of the Supplementary Information, in the above two equations (\textit{only}), $A_{ss}$ has the same units as the dimensional intracavity field, $E$, and $w_{ss}$ has the units of dimensional fast time, $\tau$. Meanwhile, $\Omega$ is the dimensional frequency (in units of inverse slow time, $t$), and
\begin{eqnarray}
\zeta &=& \frac{\beta_2 L}{\alpha \tau_s^2} - \frac{2}{3}\pi^2 \frac{w_{ss}^2}{\tau_s^2} m,\\
m &=& \frac{\gamma L}{\alpha} \left\{ \mathcal{R}(E_+ E_-)\cos(2\phi_{ss}) + \mathcal{I}(E_+ E_-)\sin(2\phi_{ss}) \right\},\ \ \ \ \ \
\end{eqnarray}
with $E_{\pm}$ being the dimensional intracavity fields.
\subsubsection{Numerical simulations}
\label{Jitter_Sims}
In this section we describe how one can compute the jitter spectrum numerically, by running many noisy LLE simulations and averaging. First of all, the quantum fluctuation terms can either be added in the temporal domain [Eq.~(\ref{time_dom})], or in the frequency domain [Eq.~(\ref{freq_dom})]. Since the LLE is solved via the SSFM, noise terms should be added to the respective step of the method, depending on whether the vacuum operators are in the time or frequency-mode representation (recall that in practice, solving the mode equations via the SSFM is the exact same process as solving the LLE). Given that the expectation values and correlations are of the same form in both representations [see Eqs.~(\ref{Vproperties}) and (\ref{Vproperties2})], we see that the noise is white and Gaussian in both cases. This means that in order to convert the quantum-mechanical equations into classical stochastic equations, we can simply replace the vacuum operators by (slow-)time-dependent noisy signals, given by complex random numbers drawn from a standard normal distribution, independently at each (slow-)time step and for each frequency mode or discretised $\tilde{\tau}$ point. Practical implementation shows better performance at very high frequencies when noise is added in the frequency domain, but in the frequency range which is physically significant, the two approaches are in agreement.

Crucially, in order to correctly solve the resulting stochastic differential equation, the noise terms must be divided by the square-root of the step size in both fast and slow times \cite{NewHope} (see also the Euler-Maruyama method).

We need to extract the soliton's position in fast time at each slow time step. There are several reasonable ways to do this, but the best and most practical method is to use the definition of the centre of mass \cite{NewHope}, calculated using the soliton intensity as the ``probability distribution'':
\begin{equation}
\delta P(\tilde{t}) = \frac{\int \tilde{\tau}|\tilde{E}|^2\ d\tilde{\tau}}{\int |\tilde{E}|^2\ d\tilde{\tau}}.
\end{equation}
For ordinary CSs, $\tilde{E}$ should be replaced by $\tilde{E}-\tilde{E}_p$, with $\tilde{E}_p$ being the background field far from the soliton. Due to the noise, one should average $\tilde{E}$ over some reasonable range of $\tilde{\tau}$ (far from the soliton) in order to accurately calculate $\tilde{E}_p$. For PDCSs, $\tilde{E}$ should be replaced by $\tilde{E}_S$. When the full bichromatically-driven PDCS LLE is used, $\tilde{E}_S$ is the field corresponding to only the central part of the spectrum (see panel (d) of Fig.~1 in the main text): the Fourier transform should be set to zero at frequencies beyond the first pair of local minima visible in the spectrum and then the inverse Fourier transform should be applied. The effect of this is to isolate the soliton, remove the driving fields and all the interference terms, and leave us with a soliton field which is essentially identical to that of the simplified model \cite{PDCS}.

Next, the dimensionless parameter $\varepsilon \equiv \sqrt{\frac{\gamma L}{\alpha} \frac{\hbar \omega_0}{t_R}}$ should be thought of as a smallness parameter which determines the magnitude of the noise. The final jitter spectrum is proportional to $\varepsilon^2$, and whenever plotted (see section \emph{Fundamental timing jitter} in the main text), it is normalised by $\varepsilon^2$. Thus the actual value of $\varepsilon$ is not important, as long as it is considerably smaller than the classical terms in the LLE. However, for practical reasons, $\varepsilon$ cannot be set \textit{too} small: we are interested in resolving the jitter in the soliton's position on a discretised (fast-)time grid. If $\varepsilon$ is taken too small, then the soliton motion can be smaller than the step size in $\tilde{\tau}$, rendering the jitter unresolvable. In practice, we adjust $\varepsilon$ so that the numerically extracted $\delta P(\tilde{t})$ is small but clear and has smooth and continuous dynamics to a good approximation.

It is also necessary to ensure that both the $\tilde{t}$ and $\tilde{\tau}$ vectors have sufficient length and resolution such that the final spectrum is no longer affected by their properties.

Once $\delta P(\tilde{t})$ is obtained for any given noisy LLE simulation (evolving the initial noise-less steady-state soliton solution from $\tilde{t}=0$ to $\tilde{t}_{max}$ and measuring the soliton position at each time step), we need to take the Fourier transform of this signal (with respect to slow time) in order to compute $\delta P(\tilde{\Omega})$. However, for reasons explained in detail in \cite{NewHope}, it is necessary to first multiply the signal with a so-called window function. Following the advice of \cite{NewHope}, this can either be an inverted quadratic which crosses zero at $\tilde{t}=0$ and $\tilde{t}_{max}$ and reaches a maximal value of one at $\tilde{t}_{max}/2$, or a half-period of a cosine function with the same values at the above-mentioned three points. We have implemented both window functions and ensured that their effect is very similar; we proceed with the cosine function which is supposed to give more accurate results \footnote{Note that processing the differences of $\delta P(\tilde{t})$ as suggested in \cite{NewHope} does not improve the spectrum further (we have checked this), and so we do not use this method.}.

After multiplication by the window function $W(\tilde{t})$, we take the Fourier transform of $\delta P(\tilde{t})W(\tilde{t})$, which gives us a good numerical representation of $\delta P(\tilde{\Omega})$. However, the fast Fourier transform algorithm does not conserve the energy of the signal, so $\delta P(\tilde{\Omega})$ must be renormalised to ensure that
\begin{equation}
\int |\delta P(\tilde{t})W(\tilde{t})|^2\ d\tilde{t} = \frac{1}{2\pi}\int |\delta P(\tilde{\Omega})|^2\ d\tilde{\Omega}.
\end{equation}
Many such noisy LLE trajectories need to be run -- we find that 500 trajectories gives excellent convergence and repeatability. We store $\delta P(\tilde{\Omega})$ from each one, then $|\delta P(\tilde{\Omega})|^2$ is computed and averaged over the 500 trajectories. Finally, it is necessary to divide the resulting spectrum by $\tilde{t}_{max}$, because it is in fact the power spectral density which is relevant for noisy signals \cite{NewHope}.

To summarise, numerically, the timing jitter spectrum is computed from
\begin{equation}
S_P = \frac{1}{\tilde{t}_{max}} \left\langle |\delta P(\tilde{\Omega})|^2 \right\rangle,
\end{equation}
where the angled brackets indicate an average over the many noisy trajectories (not to be confused with a quantum mechanical expectation value).

We have confirmed that
\begin{itemize}
\item $S_P/\varepsilon^2$ is independent of $\varepsilon$ (for reasonable values of $\varepsilon$),
\item $S_P/\varepsilon^2$ is converged with 500 trajectories and the level of noise visible on a logarithmic scale is very low,
\item $\tilde{t}_{max}$ is large enough and $\Delta \tilde{t}$ is small enough so that $S_P/\varepsilon^2$ is independent of the particular values chosen,
\item $\tilde{T}$ is large enough and $\Delta \tilde{\tau}$ is small enough so that $S_P/\varepsilon^2$ is independent of the particular values chosen,
\item the particular choise of $W(\tilde{t})$ does not affect $S_P/\varepsilon^2$ significantly.
\end{itemize}
We remark that the above observations pertain to the performance specifically of the implementation where noise is added in the frequency domain. They also hold true for the alternative implementation where noise is added in the time domain, but only at small and intermediate frequencies (which cover the physically significant range). At very large frequencies, some of these tests fail (presumably due to numerical issues), so it is advised to preferentially use the frequency-domain implementation which is more robust.
\subsection{Frequency comb linewidth}
\label{Linewidth_Method}
In this section we will briefly describe how LLE simulations incorporating phase noise arising from frequency fluctuations on the driving beams are done \cite{Frosz2006,Anderson2021}. Note that the exact same procedure is applicable regardless of the particular pump field or the model, so for simplicity, we leave off the subscripts which differentiate them (`0' for the single drive beam in the monochromatic LLE and `$\pm$' for the two pumps in the bichromatic LLE).

First, we choose a number $\epsilon\ll1$ the magnitude of which determines the level of frequency noise on the driving beam. Notice that it is assumed that we are working in dimensionless units, and recall that we are solving the LLE, using a slow-time step-size $\Delta \tilde{t}$. Phase noise $\delta\theta$ is given by the indefinite integral of frequency noise $\delta\tilde{f}$:
\begin{equation}
\delta\theta(\tilde{t}) = 2\pi\int\limits_{0}^{\tilde{t}} \delta\tilde{f}(\tilde{t}')\ d\tilde{t}',
\end{equation}
and the driving field $S$ in the LLE is then replaced by $S\exp(i\delta\theta)$. Now, $\delta\tilde{f}(\tilde{t}')$ is given by $\epsilon \times G /\sqrt{\Delta \tilde{t}}$, where $G$ is a random number drawn from a standard normal distribution (zero mean, unit standard deviation) and the division by square-root of the step size is needed for the same reason as in the previous section (namely, due to the Euler-Maruyama method). The spectrum of the input field $S\exp(i\delta\theta)$ can be computed by creating many (e.g.~500) different noisy time-dependent signals $\delta\theta$, computing the Fourier transform of $S\exp(i\delta\theta)$, taking the absolute value squared, and averaging it over the different trajectories, normalising the curve to peak at one. When this is done, the spectrum of such a noisy input field is found to be a Lorentzian with full-width-at-half-maximum (FWHM) given by $2\pi \epsilon^2$.

In order to inspect the spectrum of an individual comb line of a soliton, one has to solve the LLE adding phase noise to the driving fields as described above, and at each step in the slow-time evolution, store the relevant point of the Fourier transform of $\tilde{E}$ with respect to fast time: for example, the central mode $l=0$ corresponds to the central grid point in the vector of (fast-time) frequency $\tilde{\omega}$. Having obtained the complete time series, the spectrum is computed exactly as described in the above paragraph, except we use the comb-line time series instead of the input field itself. Such simulations reveal that the comb-line spectra are also Lorentzian, and their width is closely related to that of the driving field.

Note that it is important to pick $\tilde{t}_{max}$ to be sufficiently large so that the Lorentzian peak of the driving field is clearly resolved on a linear scale. It is also necessary to choose $\epsilon$ small enough for the soliton to survive for the entire length of the simulation (too much noise causes the soliton to disintegrate) and for the response to be linear, meaning that the spectra should be Lorentzian and not distorted and asymmetric.

\end{document}